\documentclass[twocolumn,pra,showpacs,superscriptaddress,longbibliography]{revtex4-1}   % use preprint or twocolumn
\usepackage{graphicx}
\usepackage{amsmath}
\usepackage{amssymb}
\usepackage{epstopdf}
\usepackage{dcolumn}% Align table columns on decimal point
\usepackage{bm}% bold math
\usepackage{verbatim}
\usepackage{amsfonts}
\usepackage{cancel}
\usepackage[utf8]{inputenc}
\usepackage{graphicx}
\usepackage{setspace}
\usepackage{tabularx}
\usepackage[export]{adjustbox}

\newcommand{\sspace}{$\enspace$}
\newcommand{\ssspace}{$\quad$}

\newcommand{\brakett}[3]{\left.\left\langle#1\right.\right.\|#2\|\left.\left.#3\right\rangle\right.}

\newcommand{\figref}[1]{Fig.~\ref{#1}}
\newcommand{\eqnref}[1]{Eqn.~\eqref{#1}}

\newcommand{\ak}{\alpha_{\textrm{K}}}
\newcommand{\arb}{\alpha_{\textrm{Rb}}}
\newcommand{\acs}{\alpha_{\textrm{Cs}}}

\newcommand{\polK}{42.93(2)(7)}
\newcommand{\polRb}{47.39(3)(8)}
\newcommand{\polCs}{59.39(3)(9)}
\newcommand{\polKSysOnly}{42.93(7)}
\newcommand{\polRbSysOnly}{47.39(8)}
\newcommand{\polCsSysOnly}{59.39(9)}

\newcommand{\ratRbK}{1.1040(9)}
\newcommand{\ratCsK}{1.3834(9)}
\newcommand{\ratCsRb}{1.2532(10)}

\newcommand{\Omegalab}{\Omega_{\mathrm{lab},y}}

\newcommand{\dphisepk}{\Delta\Phi_{\mathrm{sep},k}}

\newcommand{\dphisag}{\Delta\Phi_{\mathrm{sag}}}
\newcommand{\dphiaccel}{\Delta\Phi_{\mathrm{accel}}}

\newcommand{\rcs}{R_{\mathrm{Cs}}}

\newcommand{\etal}{\textit{et al.}}
\newcommand{\etalspace}{\textit{et al. }}

\newcommand{\AAA}{\mathrm{\AA}}

\newcommand{\abinit}{\textit{ab initio}}
\newcommand{\abinitspace}{\textit{ab initio} }

\begin{document}

%\title{Improved Absolute and Ratio Measurements of Ground-State Polarizabilities of Cs, Rb, and K using Atom Interferometry}
\title{Measurements of the Ground-State Polarizabilities of Cs, Rb, and K using Atom Interferometry}

\affiliation{Department of Physics, University of Arizona, Tucson, AZ 85721}
\affiliation{College of Optical Sciences, University of Arizona, Tucson, AZ 85721}
\author{Maxwell D. Gregoire}
\affiliation{Department of Physics, University of Arizona, Tucson, AZ 85721}
\author{Ivan Hromada}
\affiliation{Department of Physics, University of Arizona, Tucson, AZ 85721}
\author{William F. Holmgren}
\affiliation{Department of Physics, University of Arizona, Tucson, AZ 85721}
\author{Raisa Trubko}
\affiliation{College of Optical Sciences, University of Arizona, Tucson, AZ 85721}
\author{Alexander D. Cronin}
\affiliation{Department of Physics, University of Arizona, Tucson, AZ 85721}
\affiliation{College of Optical Sciences, University of Arizona, Tucson, AZ 85721}
\email{cronin@physics.arizona.edu}
\homepage{http://www.atomwave.org}

\date{\today}

\begin{abstract}
We measured the ground-state static electric-dipole polarizabilities of Cs, Rb, and K atoms using a three-nanograting Mach-Zehnder atom beam interferometer. Our measurements provide benchmark tests for atomic structure calculations and thus test the underlying theory used to interpret atomic parity non-conservation experiments.
We measured $\acs = 4\pi\epsilon_0 \times \polCsSysOnly \AAA^3$, $\arb = 4\pi\epsilon_0 \times \polRbSysOnly \AAA^3$, and $\ak = 4\pi\epsilon_0 \times \polKSysOnly \AAA^3$. In atomic units, these measurements are $\acs = 401.2(7)$, $\arb = 320.1(6)$, and $\ak = 290.0(5)$. We report ratios of polarizabilities $\acs/\arb = \ratCsRb$, $\acs/\ak = \ratCsK$, and $\arb/\ak = \ratRbK$ with smaller fractional uncertainty because the systematic errors for individual measurements are largely correlated. 
Since Cs atom beams have short de Broglie wavelengths, we developed measurement methods that do not require resolved atom diffraction.
Specifically, we used phase choppers to measure atomic beam velocity distributions, and we used electric field gradients to give the atom interference pattern a phase shift that depends on atomic polarizability.
\end{abstract}


\maketitle

\section{Introduction} \label{sectionIntroduction}

Measurements of static electric-dipole polarizabilities serve as benchmark tests for \textit{ab initio} calculations of electric-dipole transition matrix elements. 
These calculations require understanding quantum many-body systems with relativistic corrections, and there are many different methods that attempt to calculate these matrix elements in a reasonable amount of computing time \cite{Mitroy2010}. Testing these methods is important because these matrix elements are used to calculate several atomic properties, such as lifetimes, oscillator strengths, line strengths, van der Waals interaction potentials and associated cross sections, Feshbach resonances, and photoassociation rates.
Measuring alkali static polarizabilities as a means of testing atomic 
structure calculations has been of interest to the physics community since
Stark's pioneering measurements in 1934 \cite{Scheffers1934}. Static polarizabilities have been measured using deflection \cite{Scheffers1934,Chamberlain1963,Hall1974,Ma2015}, an E-H gradient
balance \cite{Salop1961,Molof1974a}, times-of-flight of an atomic fountain \cite{Amini2003}, and phase shifts in atomic and molecular interferometers
\cite{Ekstrom1995,Miffre2006,Holmgren2010,Berninger2007}.

We measured the static electric-dipole polarizabilities of K, Rb, and Cs atoms with 0.16\% uncertainty using a Mach-Zehnder three-grating atom interferometer \cite{Berman1997,Cronin2009} with an electric field gradient interaction region. We used the same apparatus for all three elements, so we can also report polarizability ratios with 0.08\% uncertainty because the sources of systematic uncertainty are largely correlated between our measurements of different atoms' polarizabilities. 

We compare our measurements to
\textit{ab initio} calculations of atomic polarizabilities and to polarizabilities deduced from studies of atomic lifetimes, Feshbach resonances, and photoassociation specroscopy. We also use our measurements to report the Cs $6p_{1/2}$ and $6p_{3/2}$ state lifetimes, Rb $5p_{1/2}$ and $5p_{3/2}$ state lifetimes, and K $4p_{1/2}$ and $4p_{3/2}$ state lifetimes and the associated principal electric dipole matrix elements, oscillator strengths, and line strengths. 
Then we use our measurements to report van der Waals $C_6$ coefficients,
and we combine our measurements with measurements of 
transition Stark shifts to report some excited state polarizabilities with better than 0.09\% uncertainty.

Testing Cs atomic structure calculations by measuring $\acs$ is valuable for atomic parity non-conservation (PNC) research, which places constraints on beyond-the-standard-model physics. The PNC amplitude $E_{\mathrm{PNC}}$ due to $Z^0$-mediated interactions between the Cs valence electron and the neutrons in its nucleus can be written in terms of electric dipole transition matrix elements and the nuclear weak charge parameter $Q_W$. Atomic structure calculations are needed to deduce a value of $Q_W$ from an $E_{\mathrm{PNC}}$ measurement \cite{Blundell1992,Cho1997,Derevianko2001,Porsev2009} to compare to the $Q_W$ predicted by the standard model \cite{Bouchiat1999,Dzuba2012}. Our measurement of $\acs$ tests the methods used to calculate the relevant matrix elements and provides a benchmark for the $\brakett{6s_{1/2}}{\hat{D}}{6p_{1/2}}$ matrix element, one of the terms in the expression for $E_{\mathrm{PNC}}$.

\begin{figure*}
\includegraphics[width=\linewidth,keepaspectratio]{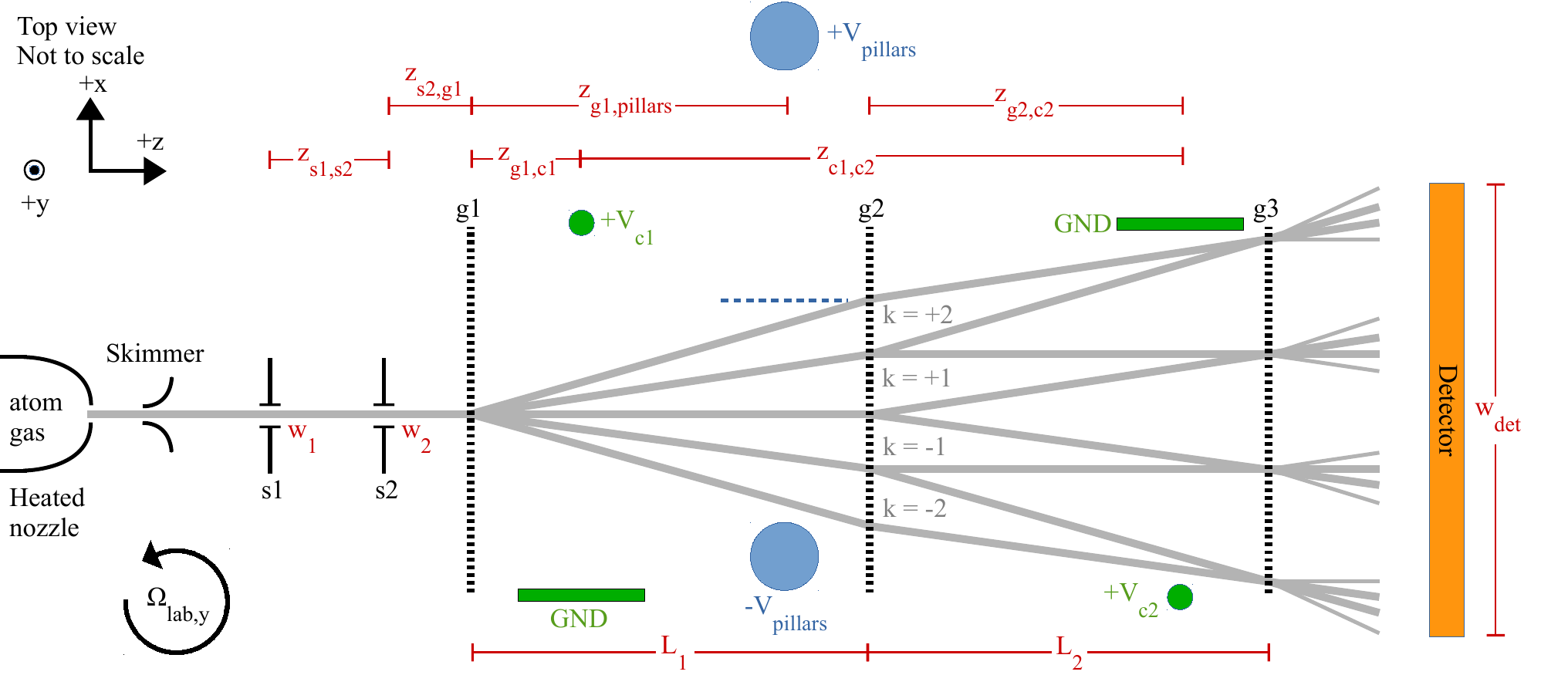}
\caption{\label{IFMDiagram}(Color online) Diagram of the Mach-Zehnder atom interferometry apparatus with phase choppers (green) and electrodes that produce an electric field gradient (blue). Dimensions are shown in red and tabulated in Table \ref{tableDimensions}.  The supersonic atom beam, shown in gray, is collimated by two slits s1 and s2 with widths $w_1$ and $w_2$ before entering the first grating. 
The nanogratings, g1, g2, and g3, are spaced longitudinally such that $L_1 = L_2$, which causes an interference pattern to form at the position of g3.
We consider the four separate interferometers in our data analysis, labeled with $k=+2,+1,-1,-2$, that form via 0th, $\pm$1st, and $\pm$2nd order diffraction from g1.
The atoms are detected by a platinum Langmuir-Taylor detector \cite{Delhuille2002}, indicated in orange.
The pair of blue circles represents oppositely-charged cylindrical electrodes (extending perpendicular to the page) that form a virtual ground plane between them. The electric field from these electrodes polarizes the atoms and thereby shifts the interference pattern's phase. 
The phase choppers are shown in green; each phase chopper is a charged wire next to a grounded plane. The geometry terms relevant to the pillars and phase choppers are displayed in \figref{EDiagram} and discussed in Section \ref{sectionPhaseShifts}.
Due to the rotation of the Earth, the lab has a rotation rate about the vertical axis of $\Omegalab = 38.88$ $\mu$rad/s that is also relevant to our analysis.  }
\end{figure*}

This is the first time that atom interferometry measurements of polarizability have been reported with smaller fractional uncertainty than the pioneering sodium polarizability measurement by Ekstrom \etalspace in 1995 \cite{Ekstrom1995}.
This is also, to our knowledge, the first time atom interferometry has been used to measure Cs polarizability.
Because it is challenging to resolve Cs atomic diffraction---our nanogratings diffract our Cs atom beams with only 20 $\mu$rad between diffraction orders---we designed an experiment 
with an electric field gradient instead of a septum electrode, such as was used in \cite{Ekstrom1995,Miffre2006}.
We also developed phase choppers \cite{Roberts2002,Roberts2004,Holmgren2011,Hromada2014} to measure our atom beams' velocity distributions instead of 
using atom diffraction to study velocity distributions, as was done in \cite{Ekstrom1995,Holmgren2010}.
These two innovations enable us to measure polarizabilities of heavy atoms such as Cs without resolving diffraction patterns. 
Without the need to resolve diffraction, we can use larger collimating slits and a wider detector to obtain data more quickly. These innovations also reduce some systematic errors that are related to beam alignment imperfections.

We improved the accuracy of our measurements compared to our previous work \cite{Holmgren2010} by redesigning the electrodes that apply phase shifts to our interferometer. 
The new configuration of electrodes, two parallel, oppositely-charged cylindrical pillars, allows us to determine the distance between the atom beam and the virtual ground plane between the pillars with reduced statistical uncertainty. We reduced systematic error by making more accurate measurements of the width of the gap between the pillars, the pillars' radii, the voltages on the pillars, and the distance between the pillars and the first diffraction grating.
Our measurements also required a sophisticated model of the apparatus, which included
interference formed by the 0th, $\pm$1st, and $\pm$2nd diffraction orders, the finite thickness and divergence of the beam, and the finite width of the detector \cite{Hromada2014}.
Because beams of Cs, Rb, and K had different velocity distributions and diffraction angles, we developed a more detailed error analysis in order to understand how those attributes affected the systematic uncertainties in polarizability measurements of different atoms.
To support our error analysis, we also developed a method to monitor and adjust the distances between nanogratings in our interferometer.

%H. Scheffers and J. Stark, Phys. Z. 35, 625 (1934)

\section{Apparatus description and error analysis} \label{sectionApparatusDescription}

\begingroup
\begin{table}
\caption{\label{tableDimensions}List of apparatus dimensions described in \figref{IFMDiagram} and \figref{EDiagram}. Dimensions with no quoted uncertainty have uncertainty much less than what would be significant to our analysis.
$a_{c1}$ and $a_{c2}$ are the closest distances between the wires and the ground planes for phase choppers 1 and 2, and $a_{\mathrm{pillars}}$ is half the width of the gap between the pillars (the closest distance between the virtual ground plane and either pillar). $L_1-L_2 = 0 \pm 30$ $\mu$m, and the uncertainty in $L_1 + L_2$ is insignificant.}
\begin{center}
\begin{tabular}{l l}
\hline\hline
$z_{\mathrm{s1,s2}}$ & 860 mm \\
$z_{\mathrm{g1,pillars}}$ & 833.5 $\pm$ 0.25 mm \\
$z_{\mathrm{g1,c1}}$ & 269.7 mm \\
$z_{\mathrm{s2,g1}}$ & 100 mm \\
$z_{\mathrm{g2,c2}}$ & 598 mm \\
$z_{\mathrm{c1,c2}}$ & 1269.3 $\pm$ 0.25 mm \\
$L_1$ & 940 mm \\
$L_2$ & 940 mm \\
$w_1$ & 30 $\pm$ 6 $\mu$m \\
$w_2$ & 40 $\pm$ 6 $\mu$m \\
$w_{\mathrm{det}}$ & 100 $\pm$ 3 $\mu$m \\ 
$a_{\mathrm{pillars}}$ & 1999.85 $\pm$ 0.5 $\mu$m \\
$R_{\mathrm{pillars}}$ & 6350 $\pm$ 0.5 $\mu$m \\
$a_{c1}$ & 986 $\pm$ 25 $\mu$m \\
$R_{c1}$ & 785.5 $\mu$m \\
$a_{c2}$ & 893 $\pm$ 25 $\mu$m \\
$R_{c2}$ & 785.5 $\mu$m \\
\hline\hline
\end{tabular}
\end{center}
\end{table}
\endgroup

A schematic diagram of the three-grating Mach-Zehnder atom beam interferometer we use to make our measurements is shown in \figref{IFMDiagram}. 
A mixture of He and Ar gas carries Cs, Rb, or K vapor through a 50 $\mu$m nozzle to generate a supersonic atom beam \cite{Scoles,Ekstrom1993}. We adjust the carrier gas composition to change the beam's average velocity: a higher percentage of Ar results in a slower beam.
The atom beam passes through two collimating slits and diffracts through three silicon nitride nanogratings, each with period $d_g = 99.90(0.5)$ nm \cite{Savas1995,Savas2003}.
The first two gratings manipulate the atoms' de Broglie waves to form a 99.90 nm period interference pattern at the position of the third grating. 
The method of observing interference fringes is described in detail in \cite{Kokorowski2001}: we scan the second grating in the $\pm x$ direction and observe the flux admitted through the third grating in order to determine the interference pattern's contrast and phase.
We measure that transmitted atomic flux with a 100 $\mu$m wide platinum wire Langmuir-Taylor detector \cite{Delhuille2002}.

In the rest of Section \ref{sectionApparatusDescription} we describe how we measure the atoms' velocity distribution and polarizability.
We measure $v_0$, the atoms' mean velocity, using phase choppers, which are charged wires parallel with the $y$ axis held parallel to grounded planes, indicated in green in \figref{IFMDiagram}.
We measure static polarizability $\alpha$ with a non-uniform electric field created by two oppositely charged cylindrical pillars parallel with the $y$ axis and indicated in blue in \figref{IFMDiagram}. The pillars' electric field shifts the interference fringe phase by an amount roughly proportional to $\alpha/v_0^2$. 
Section \ref{sectionPhaseShifts} describes how the electric field geoemtry of both the phase choppers and the pillars causes a differential phase shift. Section \ref{sectionVelocity} describes how we use the phase choppers to measure the velocity distribution, and section \ref{sectionPolarizability} describes how we use the pillars to measure $\alpha$. Section \ref{sectionDeterminingv0vr} discusses how we apply our knowledge of the velocity distribution to analyze the polarizability data taken with the pillars.

\subsection{Phase shifts with cylindrical electrodes} \label{sectionPhaseShifts}

\begin{figure}
\includegraphics[width=0.7\linewidth,keepaspectratio]{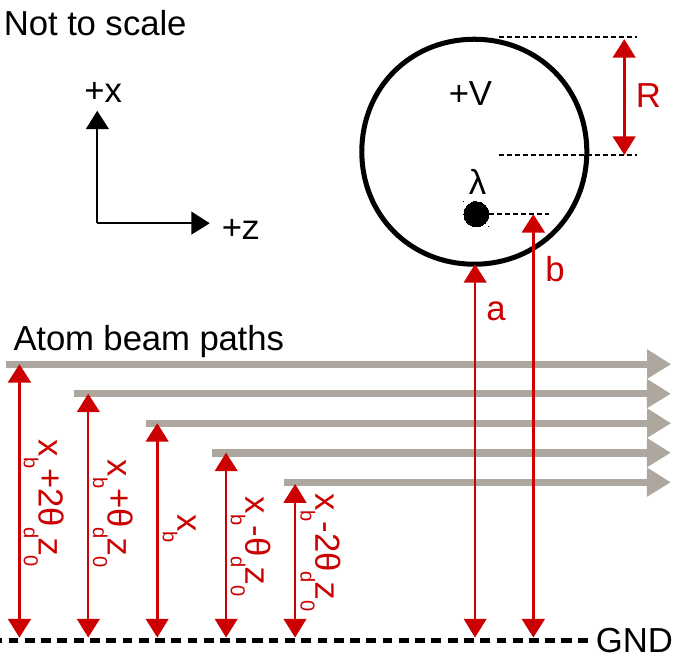}
\caption{\label{EDiagram}(Color online) Diagram showing the dimensions, shown in red, that describe the static electric fields created by the pillars and by the phase choppers. The circle represents the cross-section of a metal pillar or charged wire with radius $R$ held at voltage $V$. The GND line represents the ground plane, which may by physical (in the case of the choppers) or virtual (in the case of the pillars). $a$ is the closest distance between the pillar edge and the ground plane. The parameter $b$ is the distance between the ground plane and the effective line charge within the pillar. The atom beam center is a distance $x_b$ away from the ground plane. The different interferometer arms are separated from their neighbors by multiples of $\theta_d z_0$, where $\theta_d$ is the diffraction angle and $z_0$ is the longitudinal distance to grating g1 (in the case of the pillars and chopper c1) or grating g3 (in the case of chopper c2). $a$ and $R$ dimensions for the pillars and phase choppers are given in Table \ref{tableDimensions}.}
\end{figure}

Both the pillars and the phase choppers are described by the geometry shown in \figref{EDiagram}, and create electric fields given by
\begin{align}
	\vec{E}(x,z) = \frac{\lambda}{2\pi\epsilon_0}
	\left[	
		\frac{x-b}{(x-b)^2+z^2} - \frac{x+b}{(x+b)^2+z^2}
	\right] \hat{x} \nonumber \\
	+ 
	\left[	
		\frac{z}{(x-b)^2+z^2} - \frac{z}{(x+b)^2+z^2}
	\right] \hat{z}
	\label{EPillars}
\end{align}
where the effective line charge density
\begin{align}
	\lambda = 2\pi\epsilon_0V\ln^{-1}
	\left(
		\frac{a+R+b}{a+R-b}
	\right)
	\label{lambda}
\end{align}
exists a distance $b = a\sqrt{1+2R/a}$ away from the ground plane. The parameter $a$ represents the distance between the ground plane and the closest cylinder edge, $R$ represents the pillars' radius, and the directions $\hat{x}$ and $\hat{z}$ are shown in \figref{EDiagram}.

When atoms enter an electric field, their potential energy changes by $U_{\mathrm{Stark}} = -\frac{1}{2}\alpha|\vec{E}|^2$.
Since $U_{\mathrm{Stark}} \approx -0.1$ $\mu$eV and $E_{\mathrm{kinetic}} \approx$ 1 eV for Cs in our experiment, we can use the WKB approximation along with the Residue Theorem to compute the total phase accumulated by an atom travelling through the field.
We can also approximate that atoms travel parallel to the ground plane regardless of the angle at which they diffracted and their incident angle upon grating g1. Even though this approximation may be incorrect by up to $10^{-3}$ rad, such a discrepancy would only cause errors in the accumulated phases by factors of $10^{-6}$, which is insignificant for our experiment.
Therefore, we represent the accumulated phase along one path for a component of an atomic de Broglie wave as
\begin{align}
	\Phi(v,x) = 
	\frac{1}{\hbar v} \int_{-\infty}^{\infty} \frac{1}{2} \alpha |\vec{E}|^2 dz =	
	\frac{\lambda^2 \alpha}{\pi \epsilon_0^2 \hbar v}
	\left( \frac{b}{b^2-x_b^2} \right)
	\label{accumPhasePillars}
\end{align}
where $x_b$ is the distance between the atom's path and the ground plane.

The atoms in our beam form many interferometers, but we only need to consider the four interferometers shown in \figref{IFMDiagram}. 
Other interferometers are insignificant to our analysis because they have some combination of low contrast and low flux.
We label the four interferometers that we do consider with the index $k=+2$, $k=+1$, $k=-1$, and $k=-2$. 
The differential phase shifts for the four interferometers are
\begin{align}
	\Delta\Phi_{\vec{E},+2}(v,x_b) &= \Phi(v, x_b+2\theta_d z_0) - \Phi(v, x_b+\theta_d z_0) \nonumber \\
	\Delta\Phi_{\vec{E},+1}(v,x_b) &= \Phi(v, x_b+\theta_d z_0) - \Phi(v, x_b) \nonumber \\
	\Delta\Phi_{\vec{E},-1}(v,x_b) &= \Phi(v, x_b) - \Phi(v, x_b-\theta_d z_0) \nonumber \\
	\Delta\Phi_{\vec{E},-2}(v,x_b) &= \Phi(v, x_b-\theta_d z_0) - \Phi(v, x_b-2\theta_d z_0)
	\label{deltaPhasePillars}
\end{align}
In the above equations, $\theta_d z_0$ is the lateral separation between classical paths in the interferometer, where $\theta_d = \lambda_{\mathrm{dB}} / d_g$ is the diffraction angle and $z_0$ the distance to 
the first grating (in the case of the pillars and chopper c1) or the third grating (in the case of chopper c2).

\subsection{Velocity measurement} \label{sectionVelocity}

The atoms in the beam do not all have the same velocity, so the electric fields do not apply the same phase shifts to each diffracted atom.
We observe the average phase and contrast of an ensemble of atoms with velocity distribution $P(v)$. 
We model $P(v)$ as a Gaussian distribution
\begin{align}
	P(v) = \frac{v_r}{v_0\sqrt{2\pi}}e^{-\frac{v_r^2(v-v_0)^2}{2v_0^2}}
	\label{PvelGaussian}
\end{align}
where $v_0$ is the mean velocity and the velocity ratio $v_r = v_0/\sigma_v$ is a measure of the distribution's sharpness. It is worth noting that the velocity distribution for a supersonic atom beam is better described by a $v^3$-weighted Gaussian distribution
\cite{Berman1997}. However, either distribution can be used in our analysis to parametrize the typical high-$v_0$, high-$v_r$ velocity distributions of our atom beam without changing our polarizability result by more than 0.008\%. Since $v_0$ is the average velocity in a Gaussian but not in a $v^3$-weighted Gaussian, we use \eqnref{PvelGaussian} to simplify our discussion of the error analysis. 

%Without modeling the beam this way, we would report $\alpha$ 1.5\% too high.

\begin{figure}
\includegraphics[width=\linewidth,keepaspectratio]{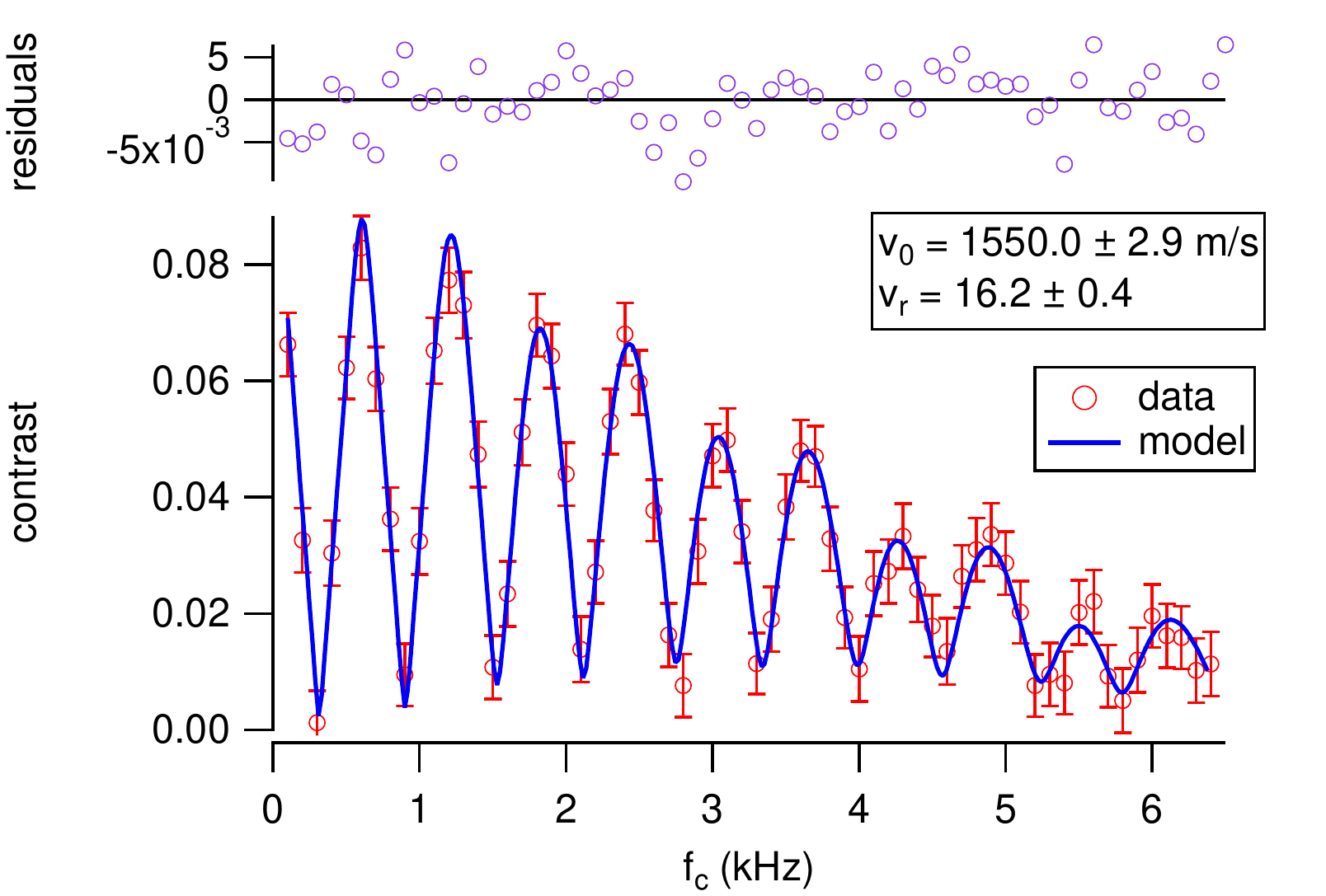}
\caption{\label{CvCFExample}(Color online) An example of a measurement of contrast $C$ vs. phase chopper frequency $f_c$ for a Cs beam. We fit a model to these data that has $v_0$ and $v_r$ as fit parameters in order to measure the velocity distribution.}
\end{figure}

To measure $v_0$ and $v_r$, we use phase choppers \cite{Holmgren2011,Hromada2014}. Each phase chopper is a charged wire about 1 mm away from a physical ground plane (see Table \ref{tableDimensions} for phase chopper dimensions). Chopper c1 is between the first two gratings and chopper c2 is a distance $z_{c1,c2} = 1269.3 \pm 0.25$ mm downstream of chopper c1, between the last two gratings (see \figref{IFMDiagram}). The voltages on the choppers' wires and the distances between the beam and the choppers' ground planes are chosen such that chopper c1 shifts the ensemble's average phase by $+\pi$ and chopper c2 shifts it by $-\pi$. 

When we pulse the choppers on and off at a frequency $f_c$, an atom may receive a net phase shift of $\pm\pi$ or $0$ depending on its velocity and the time at which it passed through the first chopper. 
Holmgren \etalspace \cite{Holmgren2011} gives an intuitive explanation of how
we measure contrast $C$ vs $f_c$ to determine $v_0$ and $v_r$. \figref{CvCFExample} shows an example of $C$ vs $f_c$ data. Hromada \etalspace \cite{Hromada2014} later improved upon Holmgren \etal's model of $C$ vs $f_c$ by considering how the thickness and divergence of the beam causes some components of the atoms' velocity distribution to not be detected. In the present work,
we expanded our analysis to include the four interferometers shown in \figref{IFMDiagram}, performed a more in-depth error analysis,
and added an additional calibration step to the measurement procedure, as we discuss next.

%Using phase choppers allows us to measure $v_0$ and $v_r$ for Cs without needing to obtain resolved diffraction. In our earlier work, we measured $v_0$ and $v_r$ by scanning the detector's $x$ position to observe the atom diffraction pattern transmission through grating g1 \cite{Holmgren2010}. However, since $\theta_d$ for Cs is so small, we would need a thinner detector wire and thinner collimating slits to resolve diffraction peaks, which would in turn reduce statistical precision.

Hromada \etalspace described how the thickness and divergence of the beam determines the likelihood for atoms of certain velocities to be detected \cite{Hromada2014}.
The thickness and divergence is defined by the finite widths of the collimating slits $w_1$ and $w_2$. The finite width of the detector $w_{\mathrm{det}}$ and the detector's offset from the beamline in the $x$ direction $\Delta x_{\mathrm{det}}$ also 
affect the probability of detecting atoms as a function of the atoms' velocities and initial positions in the apparatus.
The phase and contrast we observe with our detector is that of an ensemble of atoms with different velocities, different incident positions on grating g1, and different incident angles on grating g1.

Uncertainties in $w_1$, $w_2$, $w_{\mathrm{det}}$, and $\Delta x_{\mathrm{det}}$ are more significant for beams that are physically wider. 
In K beams, which have larger $\theta_d$ of $\approx 50$ $\mu$rad and wider velocity distributions 
($v_0 \approx 2100$ m/s, $v_r \approx 14$, and therefore $\sigma_v \approx 150$ m/s), more of the lower-velocity atoms in the distribution miss the detector. Therefore, uncertainties in the aforementioned quantities have a higher bearing on how we model the average velocity of detected atoms.
Ignoring this component of the analysis would cause a systematic increase in measured $v_0$ by 0.5\% and $v_r$ by 10\% for a typical K beam.

Modeling the four interferometers shown in \figref{IFMDiagram}, rather than only the $k=\pm1$ interferometers, also improved our understanding of how likely it is for certain velocities to be detected. For K beams with wide velocity distributions, 
we would report $v_0$ too high by about 0.5\% and $v_r$ too low by about 5\% if we included only the $k=\pm 1$ interferometers.
This is because, for such beams, a much higher proportion of atoms in the $k=\pm2$ interferometers miss the detector than in the $k=\pm1$ interferometers. 
Ignoring the $k=\pm 2$ interferometers has a significant effect on the model of the detected $P(v)$ when the detected velocity distributions for the $k=\pm1$ and $k=\pm2$ interferometers are significantly different.
Conversely, for Cs and Rb beams, we found no significant difference in results between models because most of the atoms in all interferometers were detected regardless of velocity.

Hromada \etalspace \cite{Hromada2014} also described how inequality between inter-grating distances $L_1$ and $L_2$ (see \figref{IFMDiagram}) causes systematic errors. 
When $\Delta L = L_2 - L_1$ is nonzero, the interference fringes formed at the third grating become magnified or demagnified.
We summarize this geometric magnification with the separation phase shift:
\begin{align}
	\dphisepk = \frac{2\pi}{d_g}
	\left(
		\theta_{\mathrm{inc}} + \frac{k}{2}\theta_d
	\right) \Delta L
	\label{phiSep}
\end{align}
where $k$ is the interferometer index (see \figref{IFMDiagram}) and $\theta_{\mathrm{inc}}$ is the incident angle on grating g1. 
To reduce systematic error in our $v_0$ and $v_r$ measurements, 
we measure $\Delta L$ and set it equal to zero.

\begin{figure}
\includegraphics[scale=0.5]{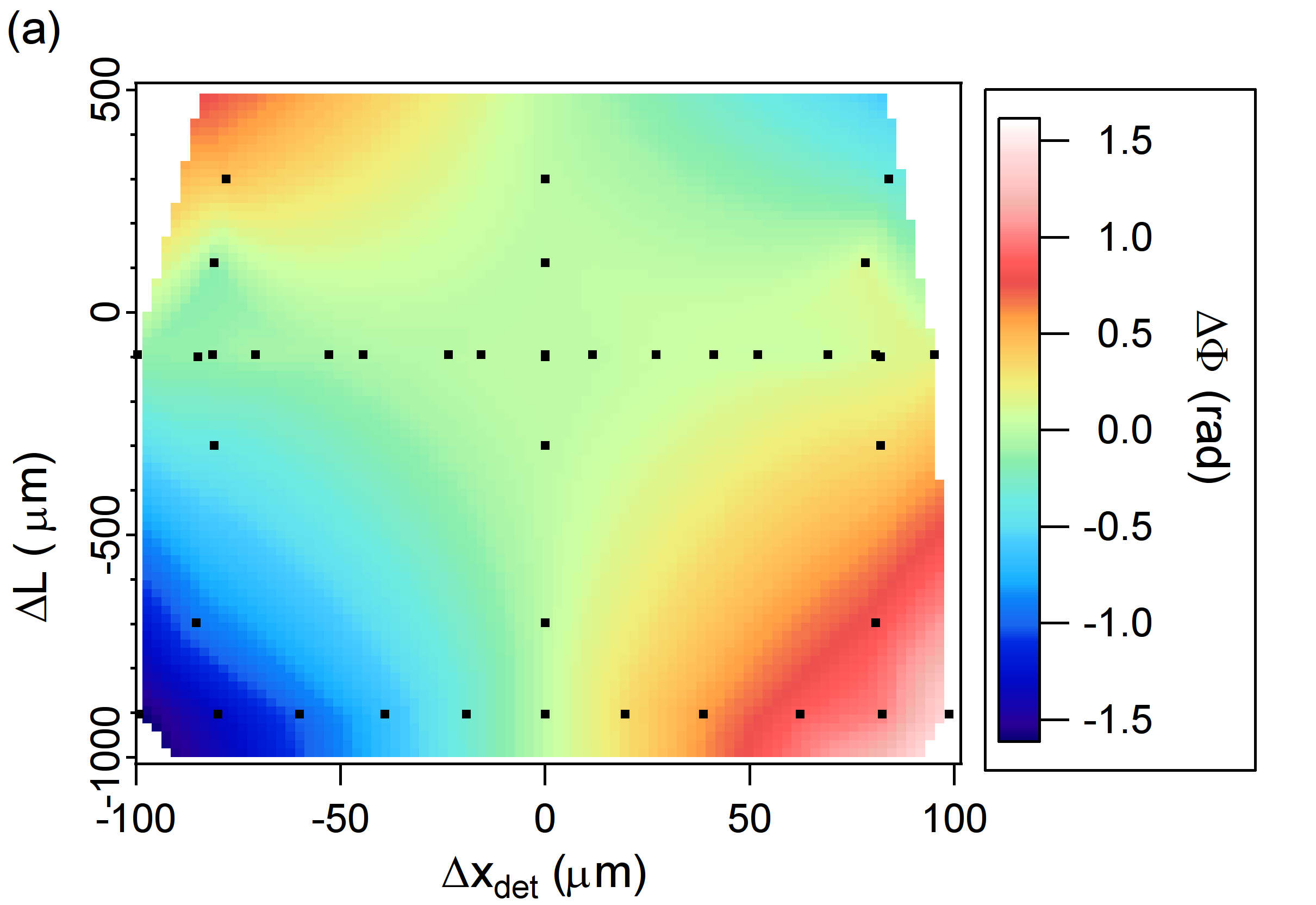}
\includegraphics[width=\linewidth,keepaspectratio]{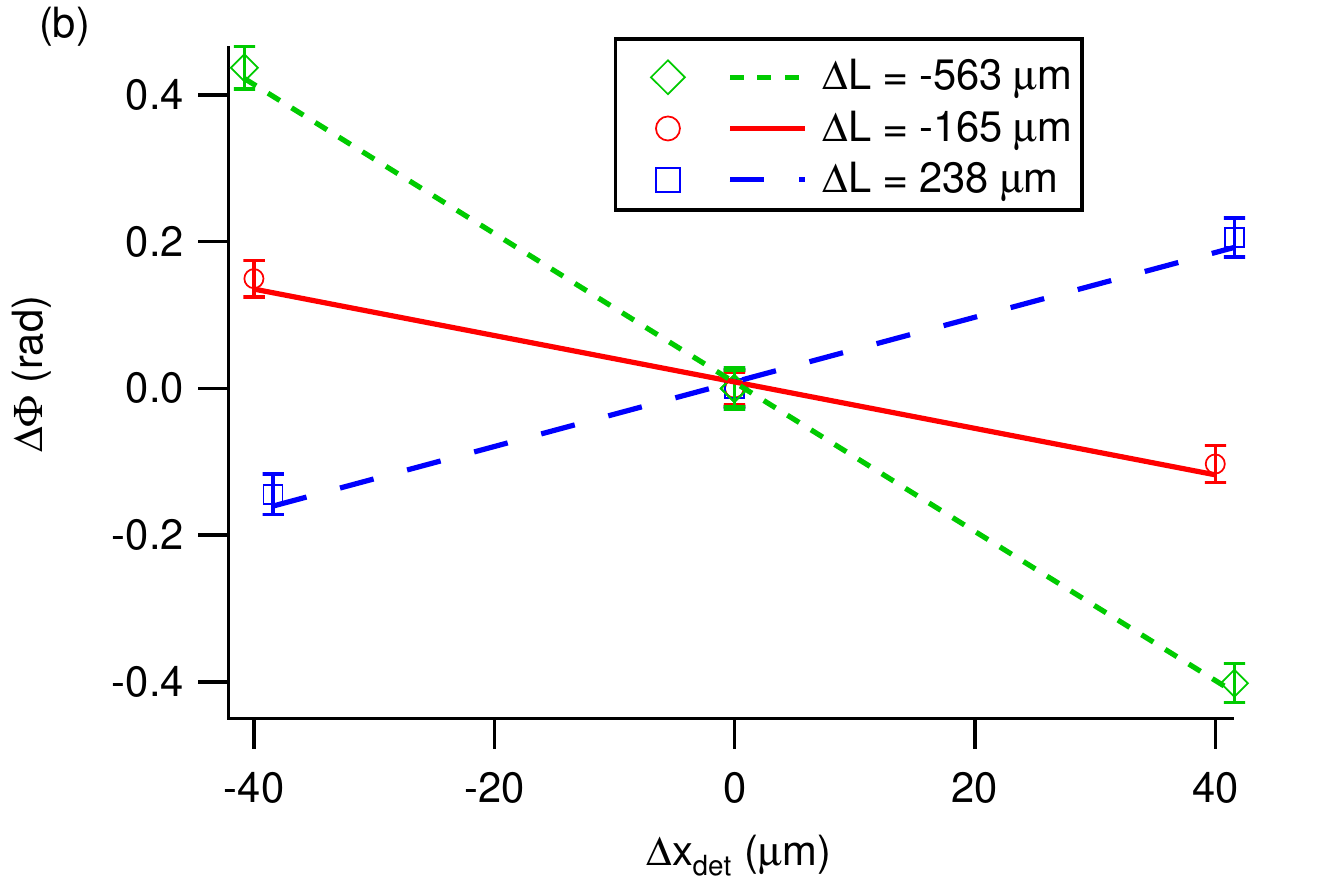}
\caption{\label{PhaseVsThings}(Color online) Two data plots showing how the phase $\Phi$ of the interference fringes changes linearly as a function of $\Delta x_{\mathrm{det}}$ and $\Delta L$.
In plot (a), the black dots represent coordinates at which data was acquired, and the colors represent contours inferred from the data.
These figures show how $\Phi$ vs $\Delta x_{\mathrm{det}}$ is a line with slope proportional to $\Delta L$. 
}
\end{figure}

\eqnref{phiSep} implies that uncertainty in $\Delta L$ is more significant for beams with larger $\theta_d$, such as K beams.
Also, because $\dphisepk$ has a component proportional to $\theta_{\mathrm{inc}}$, uncertainty in $\Delta L$ is more significant for more divergent beams.
As $|\Delta L|$ increases, uncertainties in $w_1$, $w_2$, $w_{\mathrm{det}}$, and $\Delta x_{\mathrm{det}}$ become more significant. Accordingly, we developed a method to set $\Delta L = 0$ to reduce those uncertainty contributions. 
\eqnref{phiSep} implies that interferometers on either side of the beamline receive opposite phase shifts. 
Therefore, by moving the detector in the $\pm x$ direction, we  observe linear changes in $\Phi$ as a function of $\Delta x_{\mathrm{det}}$ with slope $\mathrm{d}\Phi/\mathrm{d}x_{\mathrm{det}}$ that is proportional to $\Delta L$.
\figref{PhaseVsThings} shows data that demonstrates this effect.
We set $\Delta L$ to 0 $\pm$ 30 $\mu$m by finding the $\Delta L$ for which d$\Phi/$d$x_{\mathrm{det}} = 0$.

Since we recalibrate $\Delta L$ every day, the 30 $\mu$m uncertainty in $\Delta L$ represents a systematic error for one day's measurements and a statistical error for many days' measurements averaged together. That error will contribute toward the statistical uncertainty of $\alpha$ measurements.
The same is true for $\Delta x_{\mathrm{det}} = 0 \pm 30$ $\mu$m, which also fluctuates from day to day as we set up the apparatus.

If the interferometer grating bars are significantly non-vertical, it becomes necessary to consider the phase shift induced by the component of gravitational acceleration in the plane of the interferometer. That phase shift is given by
\begin{align}
	\dphiaccel = \frac{\pi g\sin({\theta_g})(L_1+L_2)^2}{2d_g v^2}
	\label{phiAccel}
\end{align}
where $\theta_g$ is the tilt of the grating bars with respect to vertical \cite{Greenberg2014,Trubko2015a}. See \cite{Greenberg2014} for an explanation of how we measured $\theta_g$.
Our interferometer's $|\theta_g|$ never exceeded 2.3 mrad. If we were to neglect this portion of the analysis, we would report $v_0$ incorrectly by up to 0.015\% and $v_r$ incorrectly by up to 0.25\%.

\begin{figure}
\includegraphics[width=\linewidth,keepaspectratio]{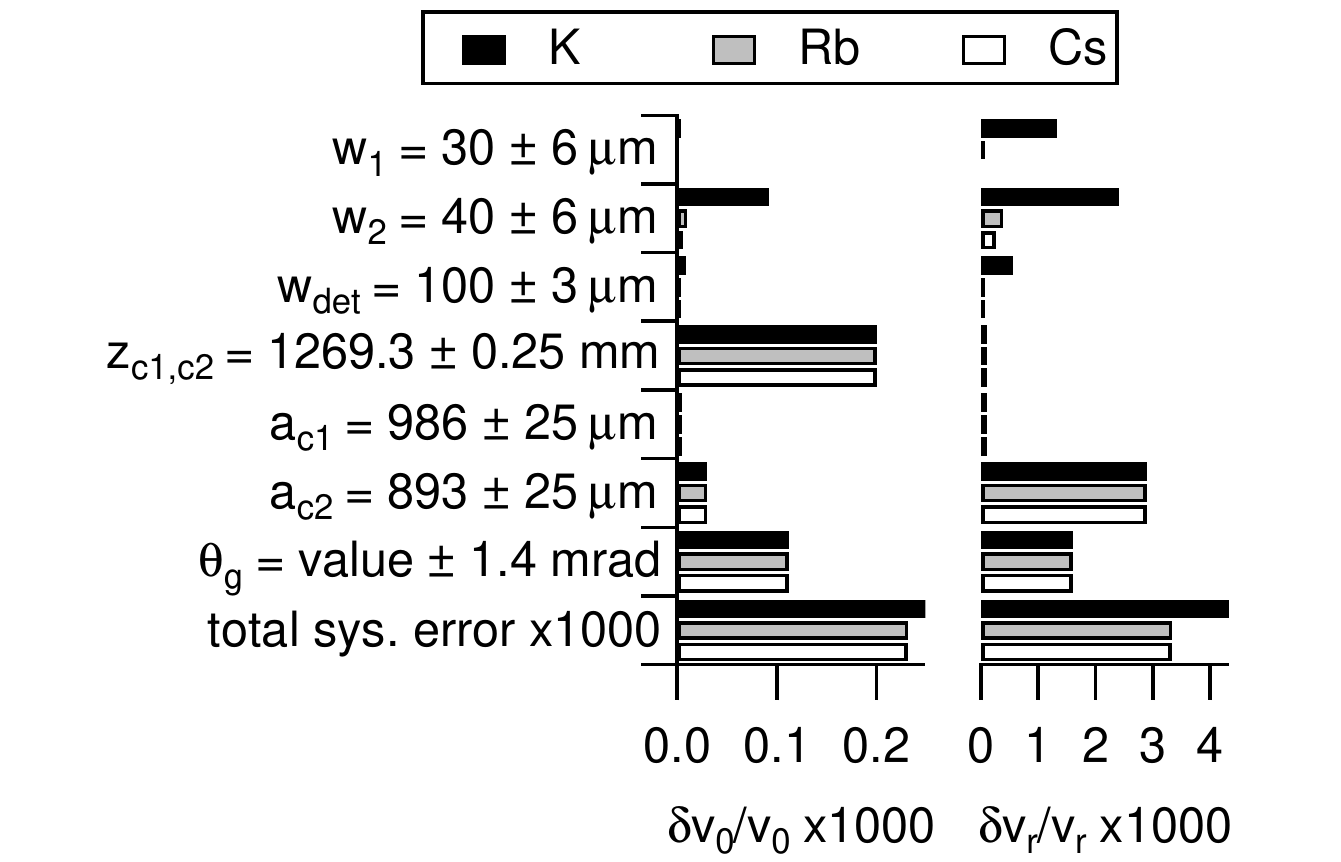}
\caption{\label{velError}(Color online) Systematic uncertainty budget for measurements of $v_0$ and $v_r$ for our Cs, Rb, and K beams. 
The total systematic error in $v_0$ and $v_r$ in turn contributes toward the total systematic uncertainty in $\alpha$ measurements (shown in \figref{polError}).
A nominal value for $\theta_g$ is not listed because $\theta_g$ changed from $-2.37 \pm 1.39$ mrad to $1.73 \pm 0.59$ mrad toward the end of the experiment.}
\end{figure}

The uncertainty budget for $v_0$ and $v_r$ measurements is displayed in \figref{velError}. 
The total statistical uncertainty in measured $v_0$ and $v_r$ is roughly 10 times larger than the total systematic uncertainty after about 15 minutes of data acquisition with the phase choppers.
Because $v_0$ and $v_r$ drift over time, typically 3\% over the course of several hours, we measure the velocity distribution twice every hour.

Gaseous alkali atoms in an atomic beam nozzle have a probability of forming homonuclear dimers that depends on the gas pressure \cite{Gordon1971} and the diameter of the nozzle hole \cite{Bergmann1978}. It is important for us to quantify the dimer mole fraction in our beam because the dimers' spatially averaged (tensor) polarizabilities are approximately 1.75 times the monomer polarizabilities \cite{Tarnovsky1993}. 
In our nozzle, the vapor pressure of alkali atoms is on the order of 1 torr at our typical running temperatures of 
160$^{\circ}$ C for Cs, 220$^{\circ}$ C for Rb, and 350$^{\circ}$ C for K.
According to data acquired by Gordon \etalspace (1971) \cite{Gordon1971} and Bergmann \etalspace (1978) \cite{Bergmann1978}, our alkali gas pressures of 1 torr should result in a dimer mole fraction well below 1\%.
Additionally, Holmgren \etalspace \cite{Holmgren2010,Holmgren2013} demonstrated how to place an upper limit on the dimer mole fraction by analyzing resolved diffraction patterns through a single nanograting and looking for peaks associated with dimer diffraction. In this work, we used very similar nozzle temperatures in our experiment as Holmgren \etalspace did in 2010. For all these reasons, we conclude that the dimer mole fraction in our beam must be less than 1\%.
\figref{polError} shows how a 4\% dimer mole fraction would lead to a significant (0.1\%) error in measured polarizability.

\subsection{Polarizability measurement} \label{sectionPolarizability}

\begin{figure}
\includegraphics[width=\linewidth,keepaspectratio]{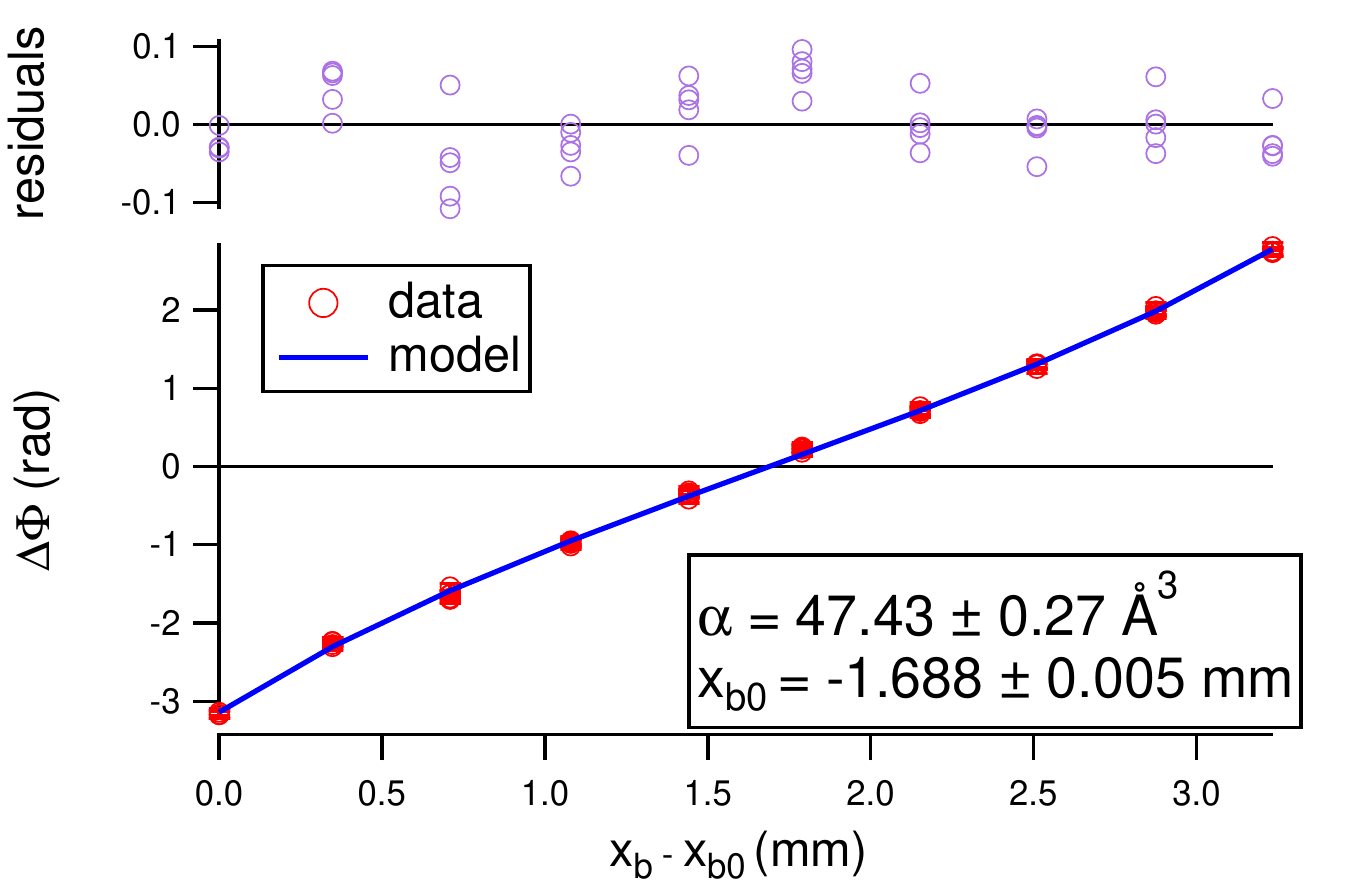}
\caption{\label{dPvMPExample}(Color online) An example of a measurement of phase shift vs $x$ position of the pillars for a Rb beam. The two fit parameters used to fit the model to these data are polarizability $\arb$ and the pillars' position at which the phase shift is null $x_{b0}$.}
\end{figure}

To measure the Cs, Rb, and K polarizability, we use two parallel, oppositely charged, $1/2$-inch-diameter, stainless-steel pillars. The pillars are mounted to a single, rigid support structure so that a 3999.7 $\pm$ 1.0 $\mu$m gap exists between them. A motor moves the support structure in the $\pm x$ direction, and a length gauge monitors the structure's $x$ position. The length gauge measures displacements of the structure with 30 nm accuracy. 
We begin a polarizability measurement with the assembly positioned such that the beam passes through the gap between the pillars near one of the edges.
We take 25 sec of data with the electric field on and 25 sec with it off.
We then move the pillars in nine 400 $\mu$m increments so that the beam approaches the other edge of the gap, taking 50 sec of data at each location.
In doing so, we
observe the phase shift $\Delta\Phi = \Phi_{\mathrm{pillars,on}} - \Phi_{\mathrm{ref}}$ applied by the pillars as a function of $x_b$ (see an example in \figref{dPvMPExample}). 
We then repeat this sequence, moving the pillars in the opposite direction in order to minimize possible systematic errors associated with travelling in a certain direction.
When the electric field is off, we observe the reference phase $\Phi_{\mathrm{ref}}$ and reference contrast $C_{\mathrm{ref}}$ given by 
\begin{align}
	C_{\mathrm{ref}}e^{i\Phi_{\mathrm{ref}}} = 
		C_0e^{i\Phi_0} \frac{1}{2} \sum_k
		\int_{v=0}^{\infty} P(v)
		e^{i\dphisag(v)} 
		dv
	\label{CPPolesRef}
\end{align}
The Sagnac phase, $\dphisag$, is a phase shift caused by the Earth's rotation and is described in \cite{Holmgren2010,Lenef1997,Jacquey2008}.
$C_0$ is the contrast that would be observed in the absence of $\dphisag(v)$, and $\Phi_0$ is an arbitrary phase constant.
When the field is on, we instead observe
\begin{align}
	& C_{\textrm{pillars,on}}e^{i\Phi_{\textrm{pillars,on}}} = \nonumber \\
		& C_0e^{i\Phi_0}		
		\frac{1}{2} \sum_k
		\int_{v=0}^{\infty} P(v)
		e^{
			i\Delta\Phi_{\vec{E},j}(v,x_b) + 
			i\dphisag(v)
		} 
		dv
	\label{CPPolesEOn}
\end{align}
We fit a model to $\Delta\Phi$ vs $x_b$, as shown in \figref{dPvMPExample}. The fit parameters of that model are the polarizability and the pillars position $x_{b0}$ for which the phase shift is zero (i.e. the location of the virtual ground plane). 

In our earlier work, we used one pillar next to a grounded plate instead of two pillars forming a virtual ground plane \cite{Holmgren2010}. We measured $x_b$ by blocking the beam with the pillar. There were significant statistical errors of a few $\mu$m associated with this procedure, and a 1 $\mu$m error would lead to a 0.1\% error in polarizability. 
Our new pillars assembly greatly reduces those statistical errors. Measuring $\Delta\Phi$ vs $x_b$ on both sides of the ground plane makes our typical 5 $\mu$m uncertainties in $x_{b0}$ add an insignificant amount of 
statistical uncertainty to the determined $\alpha$.

\begin{figure}
\includegraphics[width=\linewidth,keepaspectratio]{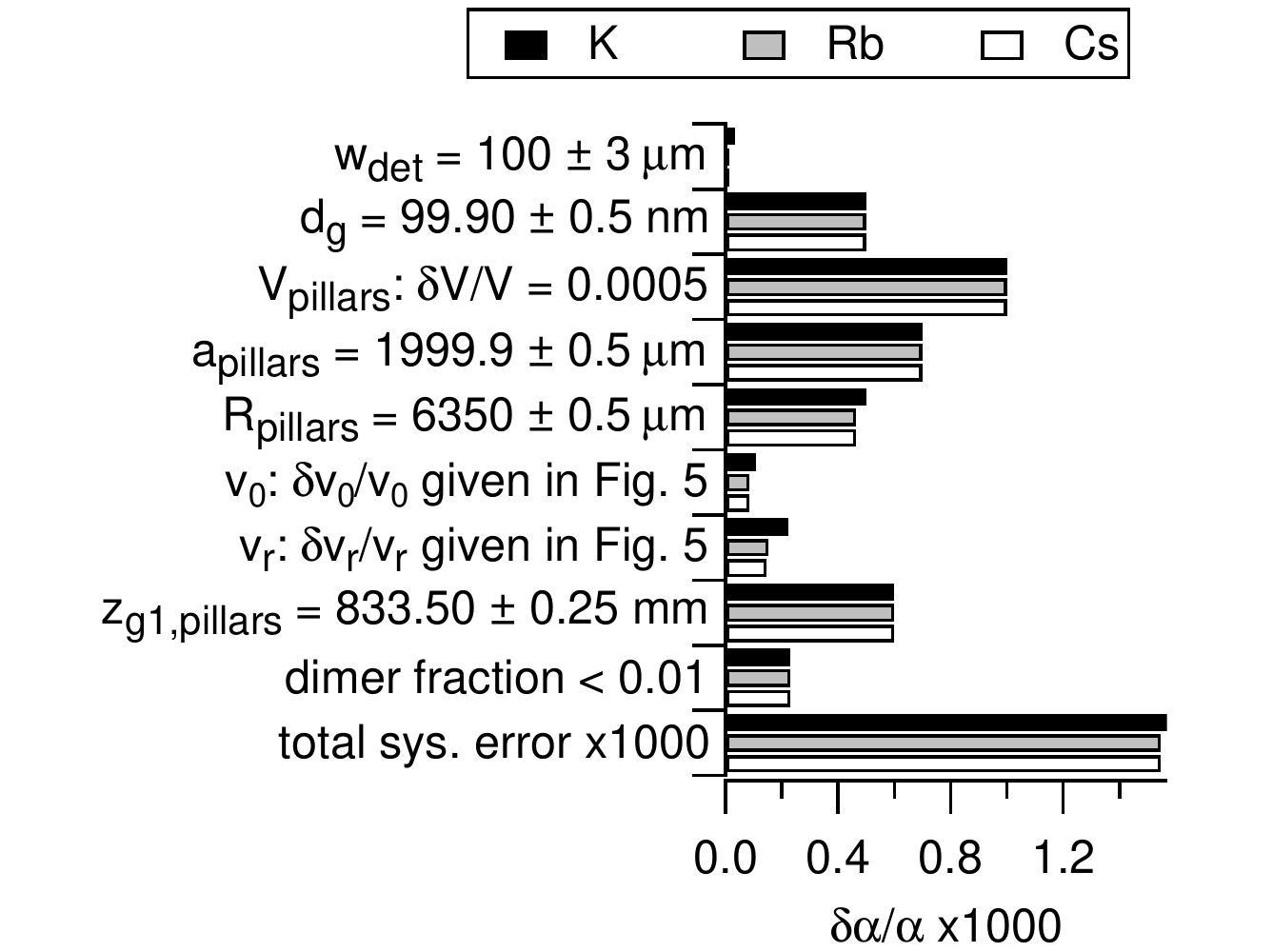}
\caption{\label{polError}(Color online) Systematic uncertainty budget for polarizability measurements for our Cs, Rb, and K beams.
The uncertainties in knowledge of $v_0$ and $v_r$ are propagated forward from \figref{velError}.
Values of $V_{\mathrm{pillars}}$ ranged from 5 kV to 7 kV, always with 0.05\% uncertainty.}
\end{figure}

The systematic uncertainty budget for our polarizability measurements is shown in \figref{polError}. In the next few paragraphs, we discuss how we measured some of the quantities in the error budget.

We reduced the uncertainty in $V_{\mathrm{pillars}}$ to 0.05\% by independently calibrating our voltage supplies. To measure $V_{\mathrm{pillars}}$, which ranged from 5 kV to 7 kV, we used a Fluke 80K-40 high voltage probe. We measured the probe's voltage divider constant, which itself depended on input voltage, using two Fluke 287 digital multimeters.

We measured $z_{g1,\mathrm{pillars}}$ to 1/4 mm accuracy. We placed rulers in the apparatus, after which three of us  would read the rulers both live and from photographs. We repeated this process for many longitudinal positions of the rulers to further reduce statistical error in the measurement. Finally, we compared the rulers we used with other rulers to verify that the ones we used were printed without significant systematic error. The value of $z_{g1,\mathrm{pillars}}$ we use in this analysis is the average of all those measurements. 

We measured the width of the gap between the pillars, $2a_{\mathrm{pillars}}$, to 1 $\mu$m accuracy by repeatedly scanning the pillars assembly across the beam and recording the positions at which each pillar blocked half of the atom beam. 
To verify that $a_{\mathrm{pillars}}$ did not change over time, we repeated this procedure many times throughout the months during which we acquired our data.

%For K beams with wide velocity distributions, we would report polarizability about 2\% lower if we didn't take into account either the $k=\pm 2$ interferometers or the beam width and divergence.

Our $\theta_g$ was always close enough to zero such that we did not need to consider $\dphiaccel$ in our polarizability data analysis. We would only need to consider $\dphiaccel$ if $|\theta_g|$ exceeded 23 mrad.
We also find that uncertainties in $w_1$ and $w_2$ each do not correspond to more than 0.004\% uncertainty in $\alpha$.

\subsection{Determining the velocity distribution during polarizability measurements} \label{sectionDeterminingv0vr}

\begingroup
\begin{table}
\caption{\label{schedule}A typical sequence of measurements during a day of data acquisition. The $+x$ direction is arbitrarily chosen---the important aspect is that we spend an equal amount of time scanning the pillars in each direction so as to minimize possible systematic errors. This sequence of eight measurements is repeated once per hour for anywhere between 8 and 36 hours. We end the data acquisition by repeating the first four measurements.}
\begin{center}
\begin{tabular}{l l}
\hline
\hline
Type of data acquired & Duration \\
\hline
contrast vs chopping freq. & 7m 5s \\
chopper c1 phase & 3m 45s \\
chopper c2 phase & 3m 45s\\
contrast vs chopping freq. & 7m 5s \\
$\Delta\Phi$ vs pillars position ($+x$ direction) \sspace & 8m 45s \\
$\Delta\Phi$ vs pillars position ($-x$ direction) & 8m 45s \\
$\Delta\Phi$ vs pillars position ($+x$ direction) & 8m 45s \\
$\Delta\Phi$ vs pillars position ($-x$ direction) & 8m 45s \\
\hline
\hline
\end{tabular}
\end{center}
\end{table}
\endgroup

A typical sequence of measurements is shown in Table \ref{schedule}.
We measure the velocity distribution twice between every four scans of the pillars across the beam, and calibrate the phase choppers between each pair of velocity measurements.

We linearly interpolate between $v_0$ and $v_r$ measurements before and after each pillars scan to estimate those quantities at the time of that scan.
Using cubic spline interpolation and Gaussian Process Regression to interpolate between $v_0$ and $v_r$ measurements changes our reported polarizabilities by no more than 0.001\%, which is small compared to our other uncertainties.

\section{Results and discussion} \label{sectionResults}

Table \ref{tableAbs} shows our measurement results for the K, Rb, and Cs atomic polarizabilities, $\ak$, $\arb$, and $\acs$.  The tabulated statistical uncertainties are the standard error of the mean for each result.  To get this statistical precision, we acquired over 90 hours of data, including 150 data sets similar to \figref{dPvMPExample} and 60 data sets similar to \figref{CvCFExample}.   The total systematic uncertainty for each measurement is also stated in Table \ref{tableAbs}, and a breakdown of the systematic uncertainty budget is summarized in \figref{polError}.  While the statistical uncertainties are typically 0.05\% for our measurements of polarizabilities, the systematic uncertainties are three to four times larger, and cause a total uncertainty of typically 0.16\% for each measurement.

We report the ratios of polarizabilities $\acs/\ak$, $\acs/\arb$, and $\arb/\ak$ in Table \ref{tableRatio}.
These ratios have uncertainties smaller than 0.08\% because we used the same apparatus for each direct measurement.
For many of the sources of systematic uncertainty summarized in \figref{polError}, an error in one of those quantities would scale each direct polarizability measurement by the same amount. These correlated uncertainties, such as electrode geometry or grating pitch, do not contribute significantly to systematic errors in our measured polarizability ratios.
However, uncertainties in $w_1$, $w_2$, $w_{\mathrm{det}}$, and $\Delta L$ affect our $\ak$, $\arb$, and $\acs$ measurements differently and therefore contribute a small amount to the final systematic uncertainties in the ratios. Even so, the ratios' systematic errors are much smaller than the statistical uncertainties. 
We discuss the value of high-precision ratios for testing atomic theories in Section \ref{sectionComparisonsWithOtherExpTheory}, and we discuss the possibility of using such ratios to improve individual measurements of polarizability in Section \ref{sectionOutlook}).

\begingroup
\begin{table}
\caption{\label{tableAbs}Direct measurements of Cs, Rb, and K static, ground-state polarizabilities.}
\begin{center}
\begin{tabular}{llll}
\hline\hline
Atom \sspace & $\alpha$(stat.)(sys.) ($\AAA^3$) \\
\hline
Cs & $\polCs$ \\
Rb & $\polRb$ \\
K  & $\polK$ \\
\hline\hline
\end{tabular}
\end{center}
\end{table}
\endgroup

\begingroup
\begin{table}
\caption{\label{tableRatio}Measured ratios of Cs, Rb, and K static, ground-state polarizabilities. The systematic uncertainties in each ratio, which arise from the fact that the systematic errors in different measurements are not perfectly correlated, are negligible compared to the statistical uncertainties.}
\begin{center}
\begin{tabular}{lll}
\hline\hline
Ratio \ssspace \ssspace & Value(stat.) \sspace & Sys. Err. \\
\hline
$\acs/\ak$  & $\ratCsK$ & $3\cdot 10^{-5}$  \\
$\acs/\arb$ & $\ratCsRb$ & $7\cdot 10^{-7}$ \\
$\arb/\ak$  & $\ratRbK$ & $2\cdot 10^{-5}$ \\
\hline\hline
\end{tabular}
\end{center}
\end{table}
\endgroup

\subsection{Comparisons with other experimental and theoretical polarizabilities} \label{sectionComparisonsWithOtherExpTheory}

\begin{figure*}
\includegraphics[width=0.60\linewidth,keepaspectratio]{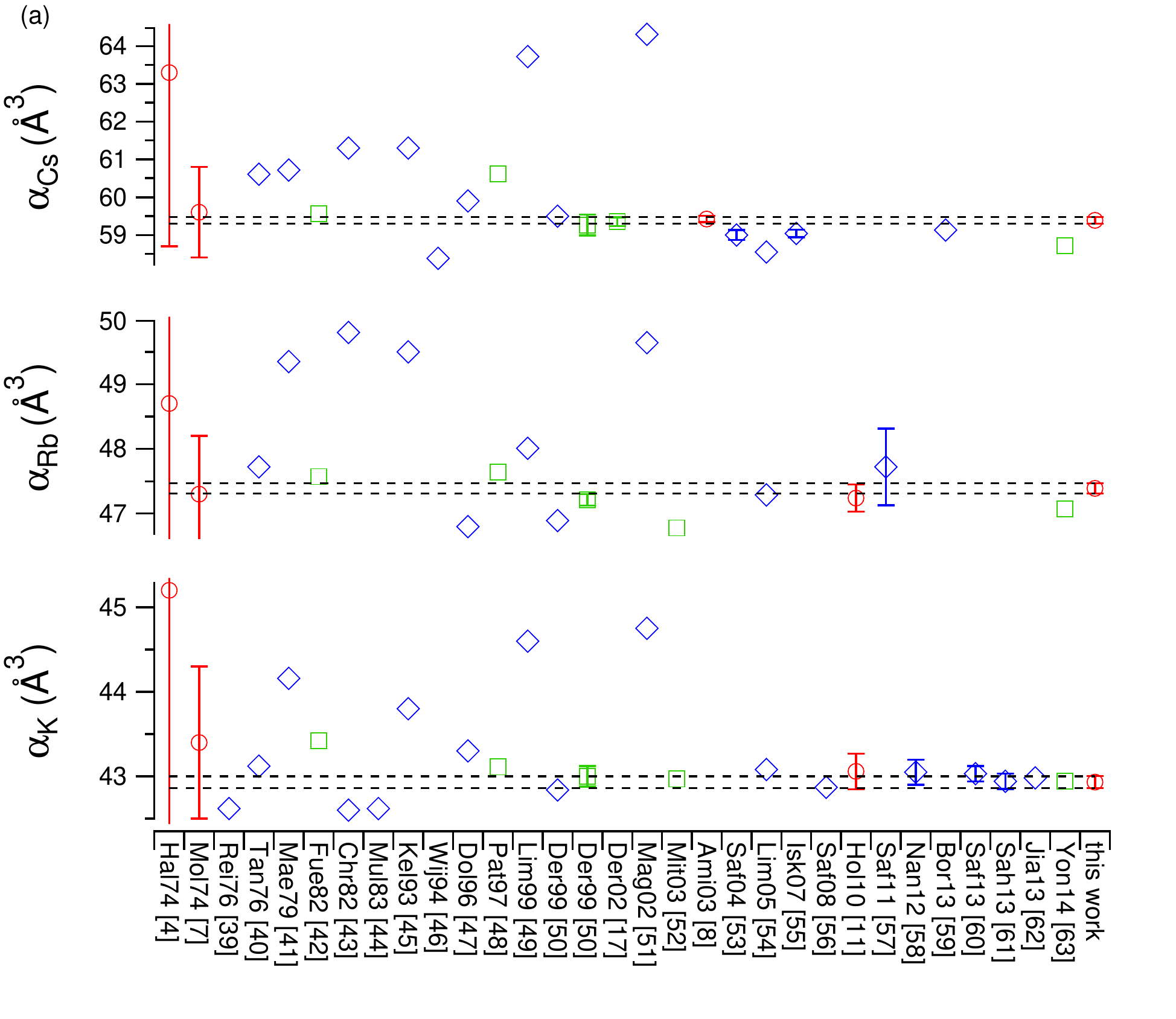}
\includegraphics[width=0.38\linewidth,keepaspectratio]{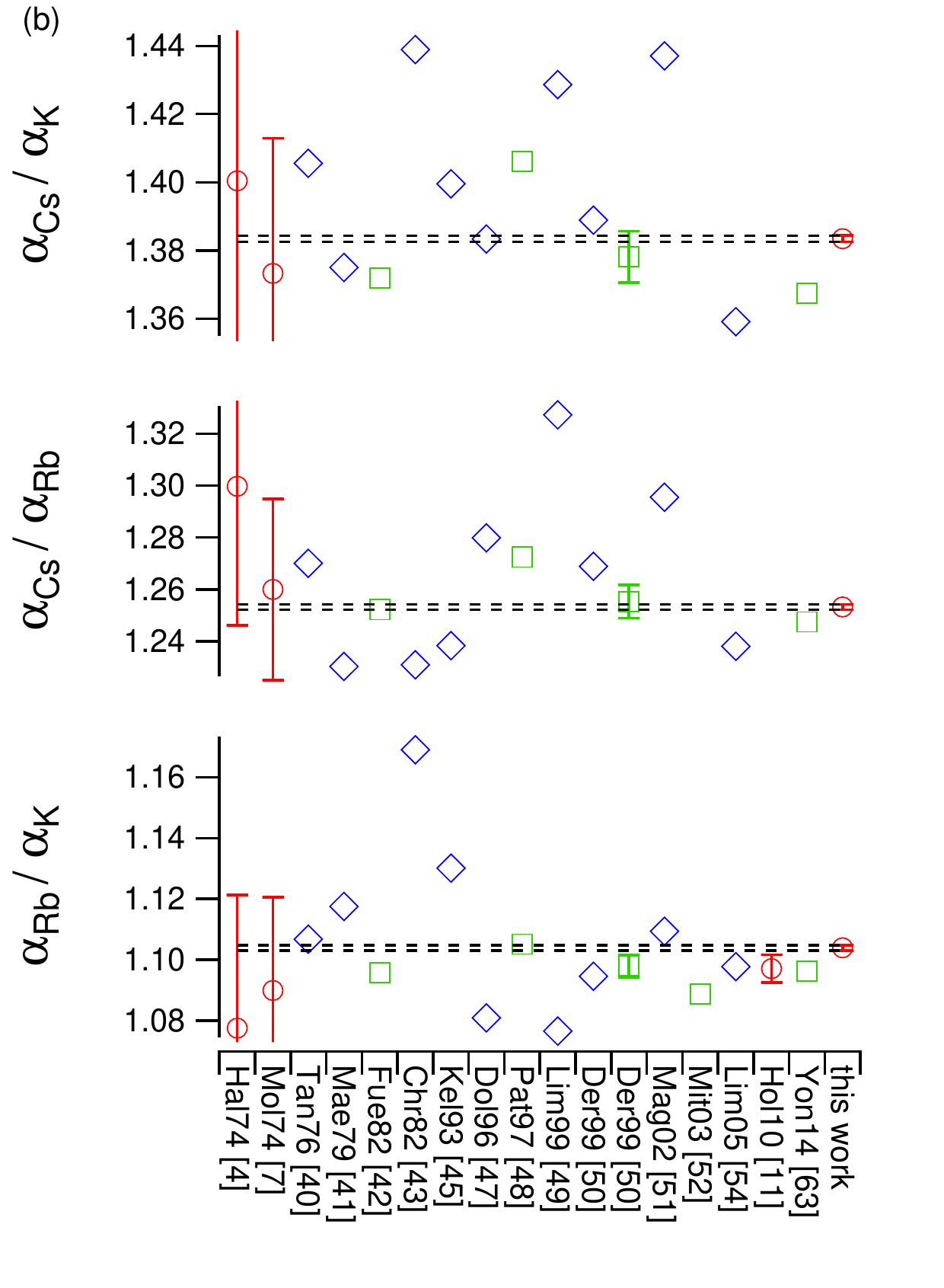}
\includegraphics[width=0.55\linewidth,keepaspectratio]{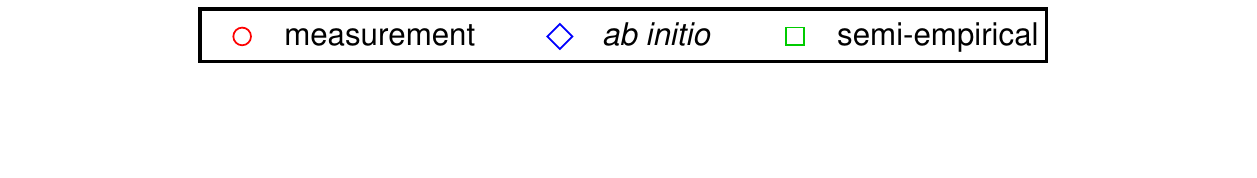}
\caption{\label{comparisons}(Color online) Our direct measurements (a) and measured ratios (b) compared with other measurements, \textit{ab initio} calculations, and semi-empirical calculations 
\cite{Hall1974,Molof1974a,Reinsch1976,Tang1976,Kutzelnigg1978,
Fuentealba1999,Christiansen1982,Muller1984,Kello1993,VanWijngaarden1994,
Dolg1996,Patil1997,Lim1999,Derevianko1998,Magnier2002,Derevianko2001,
Mitroy2003,Amini2003,Safronova2004,Lim2005,Iskrenova-Tchoukova2007,Safronova2008,
Holmgren2010,Safronova2011,Nandy2012,Borschevsky2013,Safronova2013,Sahoo2013,Jiang2013,
Safronova2013,Y.-B.2014}.
The references are represented on the x-axis by the first three letters of the first author's last name followed by the year of publication. For the semi-empirical calculations: Reference Fue82 used semi-empirical pseudopotentials \cite{Fuentealba1999}, Pat97 used experimentally-determined energy levels \cite{Patil1997}, Der99 used experimentally-determined electric dipole transition matrix elements \cite{Derevianko1998}, and Mit03 and Yon14 used semi-empirical core polarization potentials \cite{Mitroy2003,Y.-B.2014}. Values in this plot from 1999 and later are also reported in Table \ref{tableComparisons} at the end of this paper.}
\end{figure*}

\begin{figure}
\includegraphics[width=0.49\linewidth,keepaspectratio,valign=t]{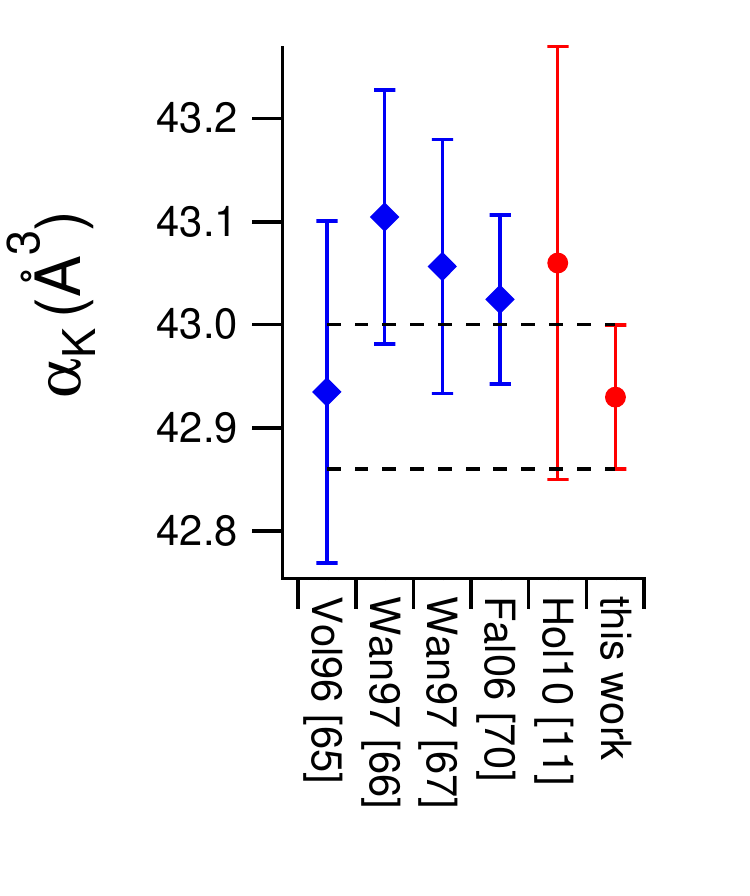}
\includegraphics[width=0.49\linewidth,keepaspectratio,valign=t]{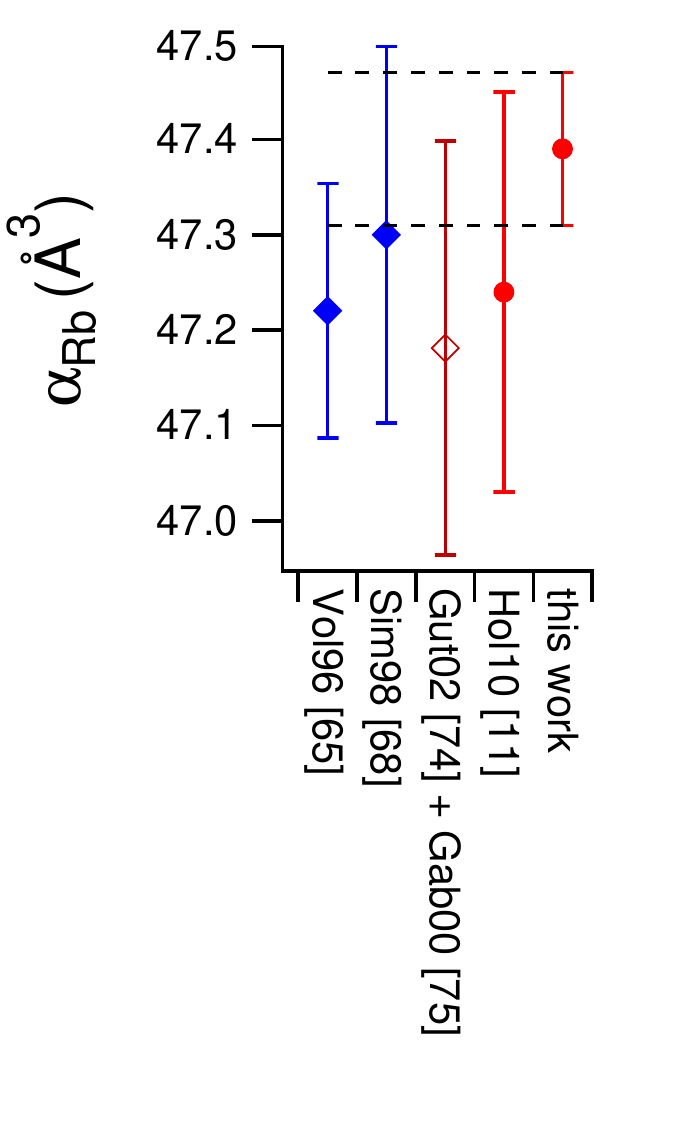}
\includegraphics[width=0.620\linewidth,keepaspectratio,valign=t]{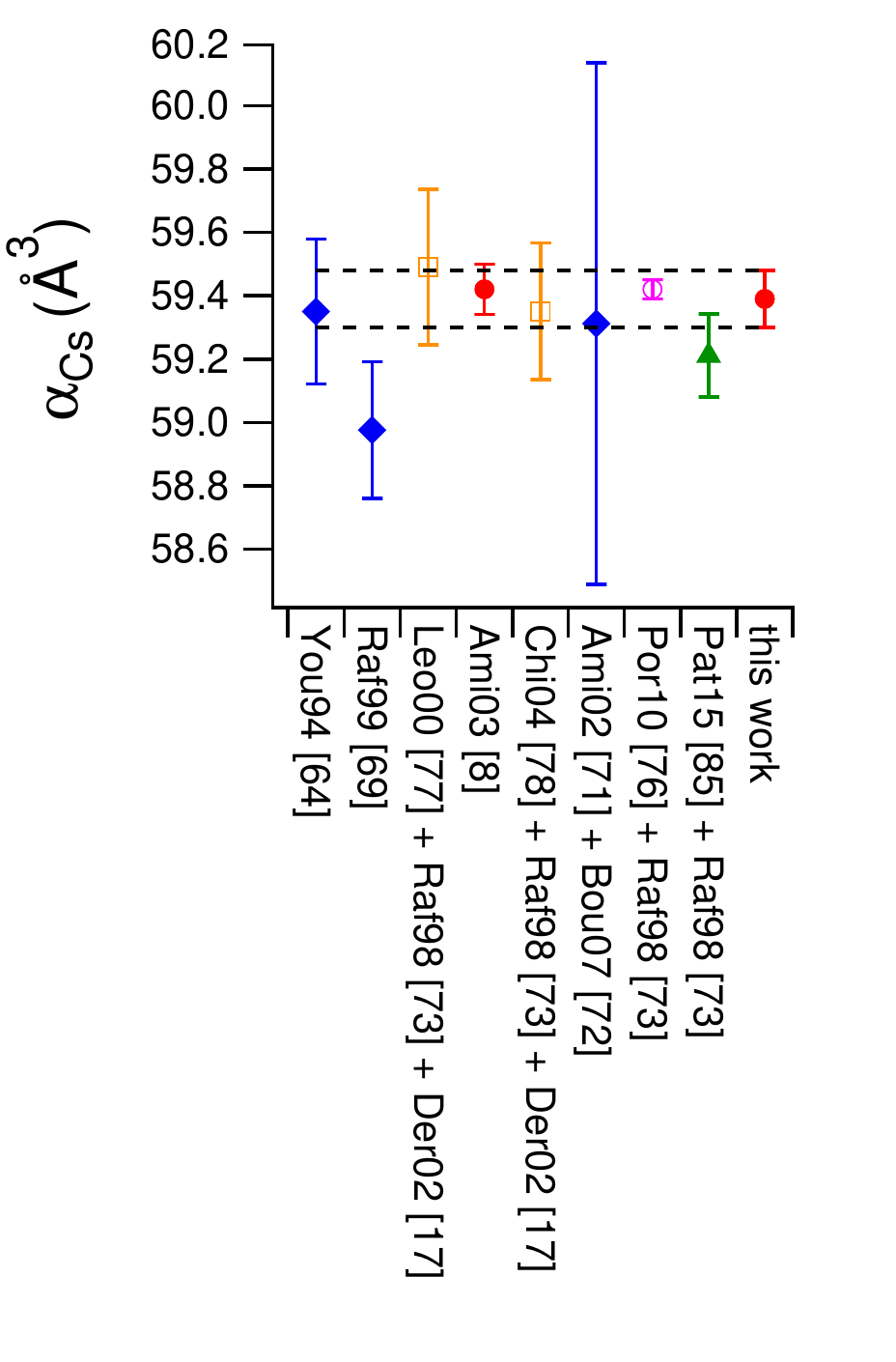}
\includegraphics[width=0.365\linewidth,keepaspectratio,valign=t]{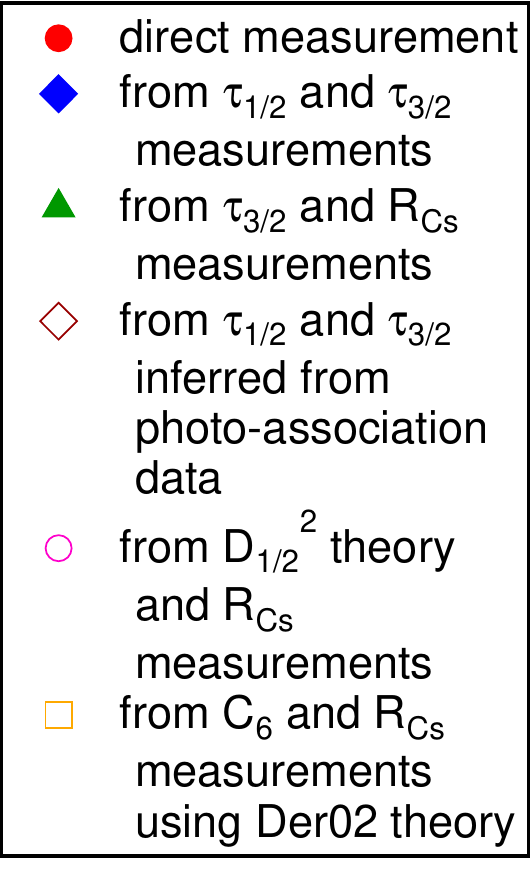}
\caption{\label{comparisonsMisc}(Color online) Comparisons of our lab's polarizability measurements (this work as well as \cite{Holmgren2010}) and Amini and Gould's $\acs$ measurement \cite{Amini2003} to polarizabilities derived from measured lifetimes and lifetime ratios,
\cite{Young1994,Volz2006,Wang1997,Wang1997a,Simsarian1998,Rafac1999,Falke2006a,Amiot2002,Bouloufa2007,Rafac1998}, 
lifetimes inferred from photo-association data
\cite{Gutterres2002,Gabbanini2000},
theoretical $D^2$ values
\cite{Porsev2010},
and van der Waals $C_6$ measurements
\cite{Leo2000,Chin2004,Derevianko2001}. Values in this plot from 1999 and later are also reported in Table \ref{tableComparisons} at the end of this paper.}
\end{figure}

\figref{comparisons} and Table \ref{tableComparisons} compare our polarizability measurements with \textit{ab initio} calculations, semi-empirical calculations, and experimental measurements subsequent to and including Molof \etal's and Hall \etal's 1974 measurements \cite{Molof1974a,Hall1974}.  First, we will discuss the comparison to previous direct measurements.   Our K and Rb polarizability measurements have 3 times smaller uncertainty than our group's previously published direct measurements of $\ak$ and $\arb$ \cite{Holmgren2010}, and 10 times smaller uncertainty than the only other direct measurements of $\ak$ and $\arb$, which were made using the E-H gradient balance technique \cite{Molof1974a} and the E gradient deflection technique  \cite{Hall1974}.  We emphasize that our new measurements are independent of the results in \cite{Holmgren2010} because although we used the same atom interferometer machine, we used a different material nanograting g1, different electrodes with different geometry, a different atom beam velocity measurement technique, a different atom beam source nozzle, and a detector with a different width. Hence, the fact that our new and more precise measurements are consistent with the measurements in \cite{Holmgren2010,Molof1974a,Hall1974} should be regarded as an independent validation of each of these previous results.

There is one other direct measurement of $\acs$ with uncertainty similar to (and slightly smaller than) ours.  Our direct $\acs$ measurement is 11 times more precise than the result reported using the E-H balance technique \cite{Molof1974a}, but 1.1 times less precise than Amini and Gould's 2003 measurement \cite{Amini2003} that was made using an atomic fountain apparatus.
To our knowledge, Amini and Gould's work is the only polarizability measurement to date that has been accomplished using an atomic fountain, and it produced a remarkable improvement in precision by a factor of 15 as compared to the only previous direct measurements of $\acs$ \cite{Molof1974a,Hall1974}.  Furthermore, $\acs$ measurements can test some of the atomic structure theory that is used to interpret atomic parity non-conservation experiments as a way of constraining physics beyond the standard model \cite{Blundell1992,Cho1997,Derevianko2001,Porsev2009}. Thus, it is particularly important to validate this $\acs$ result in \cite{Amini2003}.  We find that our $\acs$ measurement is consistent with Amini and Gould's.  Our result $\acs = \polCsSysOnly \AAA^3$ deviates from their result of $\acs = 59.42(8) \AAA^3$ by 0.03 $\AAA^3$, which is insignificant.  Comparing our atom interferometer result with their fountain result serves as a cross-check for both methods.
Both measurements also agree with $\acs$ values inferred from the atomic structure calculations by Derevianko and Porsev (2002) \cite{Derevianko2001} and Derevianko \etalspace (1999) \cite{Derevianko1998} for PNC analysis.

Most theoretical predictions for $\ak$, $\arb$, and $\acs$ deviate from each other and from our measurements significantly.   Out of 28 sets of theoretical predictions shown in \figref{comparisons}, only ten sets of predictions \cite{Derevianko1998,Derevianko2001,Iskrenova-Tchoukova2007,Safronova2008,Safronova2011,
Nandy2012,Jiang2013,Sahoo2013,
Safronova2013,Borschevsky2013} are consistent with our results within 3$\sigma$ (where $\sigma$ is the standard deviation of our measurement). Furthermore, the semi-empirical $\ak$, $\arb$, and $\acs$ values calculated in 1999 by Derevianko \etalspace \cite{Derevianko1998} are the only predictions that match all three of our own $\ak$, $\arb$, and $\acs$ measurements
to within 3$\sigma$. These predictions \cite{Derevianko1998} were made using measured lifetimes and energies, and Derevianko and Porsev's later $\acs$ prediction \cite{Derevianko2001} was made using a measured van der Waals $C_6$ coefficient.  This is an important point because there are now additional data on lifetimes, van der Waals $C_6$ measurements, and line strength ratios that can inform new semi-empirical predictions for polarizabilities that we discuss in Section \ref{sectionComparisonsWithDerived} and \figref{comparisonsMisc}.

\figref{comparisons} (b) compares our measurements of atomic polarizability ratios to other theoretical, semi-empirical, and experimental reports for these ratios.  The values we measured for $\acs/\ak$, $\acs/\arb$, and $\arb/\ak$ are consistent with all of the previous experimental measurements of these ratios, given the larger uncertainties associated with previous measurements.  Comparing theoretical predictions to our measured polarizability ratios serves as a different way to test the theoretical predictions.  Since the fractional uncertainties on our measured ratios are smaller than those of our absolute measurements, our ratios serve as a more precise test for theoretical works that predict $\alpha$ values for multiple alkali atoms.

\subsection{Comparisons with polarizabilities derived from other quantities} \label{sectionComparisonsWithDerived}

Static polarizabilities can be related to electric dipole transition matrix elements, state lifetimes, oscillator strengths, and van der Waals coefficients. We will describe those relations and compare our $\alpha$ measurements to $\alpha$ values derived from recent calculations and high-precision measurements of those quantities. Those comparisons are shown in \figref{comparisonsMisc} and Table \ref{tableComparisons}.

\begingroup
\begin{table}
\caption{\label{tableOmegaRes}We use the following residual polarizabilities $\alpha_r$ and matrix element ratios $R = D_{3/2}^2/D_{1/2}^2$.
The sources for each quantity are cited next to the values in the table.}
\begin{center}
\begin{tabular}{llll}
\hline\hline
Atom$\quad$~ & $\alpha_r$ ($\AAA^3$) \quad\quad\quad\quad\quad & $R$ \\
\hline
Cs & 2.481(16) \cite{Derevianko2001} & 1.9809(9) & \cite{Rafac1998} \\
Rb & 1.562(89) \cite{Safronova2006} & 1.996(4) & \cite{Volz2006} \\ %\cite{Leonard2015}
%Rb & 1.562(89) \cite{Safronova2006} & 1.99219(3) & \cite{Leonard2015} \\
K  & 0.925(45) \cite{Safronova2006} & 2.000(4) & \cite{Holmgren2012} \\
\hline\hline
\end{tabular}
\end{center}
\end{table}
\endgroup

The polarizability (in volume units) of an atom in state $i$ can be written in terms of Einstein A coefficients $A_{ij}$ as
\begin{align}
	\alpha_i = \frac{c^3}{2} \sum_{j\neq i} 
	\frac{A_{ij}}{\omega_{ij}^4} \frac{g_j}{g_i}
	+ \alpha_r
	\label{polFromLifetimes}
\end{align}
where $\omega_{ik}$ is the transition frequency between states $i$ and $k$, and $g_n = 2J_n+1$ is the degeneracy factor for state $n$. In our case, state $i$ is the ground state. The residual polarizability $\alpha_r$ includes terms not explicitly included in the sum, the polarizability of the core electrons, and a correction accounting for correlations between core and valence electrons as described in several references 
\cite{Derevianko1998, Derevianko2001, Safronova2006}.  We will explicitly sum over the principal transitions from $ns_{1/2}$ to $np_{1/2}$ and $np_{3/2}$, where $n=6$ for Cs, $n=5$ for Rb, and $n=4$ for K, and we will include the other transitions in $\alpha_r$. We abbreviate the lifetimes associated with the principal transitions to $\tau_{1/2}$ and $\tau_{3/2}$. In our calculations, we use the transition wavelengths $\omega_{ik}$ from references
\cite{Gerginov2005, Gerginov2006, Falke2006, Johansson1961}
and the $\alpha_r$ values indicated in Table \ref{tableOmegaRes}.
\figref{comparisonsMisc} shows polarizabilities calculated using measurements of $\tau_{1/2}$ and $\tau_{3/2}$
\cite{Young1994,Rafac1999,Bouloufa2007,Falke2006a,Volz2006,Simsarian1998,Wang1997}.
\figref{comparisonsMisc} also shows $\acs$ calculated from
values of $\tau_{1/2,\mathrm{Rb}}$ and $\tau_{3/2,\mathrm{Rb}}$ inferred in 2002 by Gutterres \etalspace from photo-association data taken in 2000 by Gabbanini \etalspace \cite{Gabbanini2000,Gutterres2002}.

We can use Patterson \etal's 2015 measurement of $\tau_{3/2,\mathrm{Cs}}$ \cite{Patterson2015} along with a measurement of the ratio of principal transition matrix elements to report $\acs$.
Rafac and Tanner measured the ratio of Cs electric dipole transition matrix elements \cite{Rafac1998}
\begin{align}
	\rcs = \frac
	{\left|\brakett{6s_{1/2}}{\hat{D}}{6p_{3/2}}\right|^2}
	{\left|\brakett{6s_{1/2}}{\hat{D}}{6p_{1/2}}\right|^2}
	\label{polFromLifetimes}
\end{align}
which is related to the ratio of lifetimes
\begin{align}
	\frac{\tau_{1/2}}{\tau_{3/2}} = \frac{\rcs}{2} \left( \frac{\omega_{3/2}}{\omega_{1/2}} \right)^3
	\label{RafacRLifetimes}
\end{align}

\begingroup
\begin{table}
\caption{\label{tableMisc}Matrix elements, lifetimes, oscillator strengths, line strengths, and van der Waals $C_6$ coefficients calculated from our polarizability measurements.
We used $R$ values from Table \ref{tableOmegaRes}. The matrix elements, line strengths, and $C_6$ coefficient are expressed in atomic units, while the lifetimes are expressed in SI units.
$\delta_{\alpha}$, $\delta_{R}$, and $\delta_{\alpha_r}$ represent the uncertainties in the values due to uncertainty in $\alpha$, $R$, and $\alpha_r$, respectively. $\delta_{\mathrm{tot}}$ is the total uncertainty in the value. (-) represents a nonzero uncertainty of less than 0.5.}
\begin{center}
\begin{tabular}{lllllll}
\hline\hline
Quantity & Atom & Value & $\delta_{\alpha}$ & $\delta_{R}$ & $\delta_{\alpha_r}$ & $\delta_{\mathrm{tot}}$ \\
\hline
$D_{1/2}$ 		& Cs & 4.508 & (4) & (1) & (1) & (4) \\
 					& Rb & 4.239 & (4) & (3) & (4) & (6) \\
 					& K  & 4.101 & (3) & (3) & (2) & (5) \\ \hline
$D_{3/2}$ 		& Cs & 6.345 & (5) & (-) & (1) & (5) \\
 					& Rb & 5.989 & (5) & (2) & (6) & (8) \\
 					& K  & 5.800 & (5) & (2) & (3) & (6) \\ \hline
$\tau_{1/2}$ (ns) 	& Cs & 34.77 & (5) & (1) & (1) & (6) \\
 					& Rb & 27.60 & (5) & (4) & (5) & (8) \\
 					& K  & 26.81 & (4) & (4) & (3) & (6) \\ \hline
$\tau_{3/2}$ (ns) 	& Cs & 30.37 & (5) & (-) & (1) & (5) \\
 					& Rb & 26.14 & (5) & (2) & (5) & (7) \\
 					& K  & 26.45 & (4) & (2) & (3) & (6) \\ \hline
$f_{1/2}$ 			& Cs & 0.3450 & (5) & (1) & (1) & (6) \\
 					& Rb & 0.3433 & (6) & (5) & (7) & (10) \\
 					& K  & 0.3317 & (6) & (4) & (4) & (8) \\ \hline
$f_{3/2}$ 			& Cs & 0.7174 & (11) & (1) & (2) & (12) \\
 					& Rb & 0.6982 & (12) & (5) & (14) & (19) \\
 					& K  & 0.6665 & (11) & (4) & (7) & (14) \\ \hline
$S_{1/2}$ 			& Cs & 20.32 & (3) & (1) & (1) & (3) \\
 					& Rb & 17.97 & (3) & (3) & (3) & (5) \\
 					& K  & 16.82 & (3) & (2) & (2) & (4) \\ \hline
$S_{3/2}$ 			& Cs & 40.26 & (6) & (1) & (1) & (6) \\
 					& Rb & 35.87 & (6) & (3) & (7) & (10) \\
 					& K  & 33.64 & (6) & (2) & (4) & (7) \\ \hline
$C_6$ 				& Cs & 6879 & (20) & (-) & (7) & (21) \\ 
					& Rb & 4719 & (15) & (-) & (26) & (30) \\
 					& K  & 3884 & (13) & (-) & (14) & (19) \\ \hline
\hline
\end{tabular}
\end{center}
\end{table}
\endgroup

\begingroup
\begin{table}
\caption{\label{tableAbs}Excited state polarizabilities $\alpha_{np_{1/2}}$, where $n = 6$ for Cs, $n = 5$ for Rb, and $n = 4$ for K. The values were calculated using our measurements and $\alpha_{np_{1/2}} - \alpha_{ns_{1/2}}$ measurements \cite{Hunter1991,Miller1994}.}
\begin{center}
\begin{tabular}{llll}
\hline\hline
Atom \sspace & $\alpha_{np_{1/2}}$ ($\AAA^3$) \\
\hline
Cs & 196.81(9) \\
Rb & 120.33(8) \\
K  & 89.92(7) \\
\hline\hline
\end{tabular}
\end{center}
\end{table}
\endgroup

\begingroup
\begin{table*}
\caption{\label{tableComparisons} Comparisons of directly measured, semi-empirical, and \abinitspace polarizabilities from 1999 and later. These values are also plotted in \figref{comparisons} and \figref{comparisonsMisc}.  Lines with $\alpha_r$ in the Method column were calculated using residual polarizabilities $\alpha_r$ in Table \ref{tableOmegaRes} \cite{Derevianko2001,Safronova2006}.  Ratios of polarizabilities $\arb/\ak$ = 1.097(5) from \cite{Holmgren2010} and $\acs/\arb$ = \ratCsRb, $\acs/\ak$ = \ratCsK, $\arb/\ak$ = \ratRbK \. from this work have been reported with smaller fractional uncertainties than the direct measurements listed in this table. For the \abinitspace method acronyms: CCSD indicates the coupled-cluster approach with single and double excitations of the wavefunctions. The prefix R stands for ``relativistic" and the prefix L stands for ``linearized". The suffix T indicates that the authors used perturbative triple excitations of the wavefunctions. See \cite{Mitroy2010} for explanations of these various theoretical methods.}
\begin{center}
\begin{tabular}{l l l l l}
\hline\hline
Reference(s) & Method & $\acs$ ($\AAA^3$) \,\,\,\,& $\arb$ ($\AAA^3$) \,\,\,\,& $\ak$ ($\AAA^3$) \\
\hline
%You94 \cite{Young1994} & $\tau_{1/2}$, $\tau_{3/2}$ meas. & 59.35(23) &  &  \\
%Vol96 \cite{Volz2006} & $\tau_{1/2}$, $\tau_{3/2}$ meas. &  & 42.22(13) & 42.94(17) \\
Raf99 \cite{Rafac1999,Derevianko2001}  & $\tau_{1/2}$, $\tau_{3/2}$ meas. + $\alpha_r$& 58.97(22) &  &  \\
Der99 \cite{Derevianko1998} & \abinit, RLCCSD & 59.50 & 46.89 & 42.84 \\ % verified
Der99 \cite{Derevianko1998} & semi-empirical & 59.26(28) & 47.21(9) & 43.00(12) \\
Leo00 \cite{Leo2000,Rafac1998,Derevianko2001}  & $C_6$ meas. + $\rcs$ meas. + thry + $\alpha_r$ \,\,& 59.49(25) & & \\
Gut02 \cite{Gutterres2002,Gabbanini2000,Safronova2006} \,\, & $\tau_{1/2}$, $\tau_{3/2}$ from PA data + $\alpha_r$ & & 47.18(22) & \\
Der02 \cite{Derevianko2001} & semi-empirical & 59.35(12) & & \\
Mag02 \cite{Magnier2002} & \abinit & 64.31 & 49.64 & 44.75 \\
Ami03 \cite{Amini2003} & direct $\acs$ meas. & 59.42(8) & & \\
Mit03 \cite{Mitroy2003} & semi-empirical & & 46.78 & 42.97 \\ % verified
Chi04 \cite{Chin2004,Rafac1998,Derevianko2001} & $C_6$ meas. + $\rcs$ meas. + thry + $\alpha_r$ & 59.35(22) & & \\
Saf04 \cite{Safronova2004} & \abinit & 59.00(13) & & \\
Lim05 \cite{Lim2005} & \abinit, RCCSDT & 58.55 & 47.29 & 43.08 \\ % verified
Fal06 \cite{Falke2006a,Safronova2006}  & $\tau_{1/2}$, $\tau_{3/2}$ meas. + $\alpha_r$& & & 43.02(8) \\
Bou07 \cite{Bouloufa2007,Derevianko2001} & $\tau_{1/2}$, $\tau_{3/2}$ meas. + $\alpha_r$ & 59.31(82) & & \\
Isk07 \cite{Iskrenova-Tchoukova2007} & \abinit, RLCCSDT & 59.04(10) & & \\ % verified
Saf08 \cite{Safronova2008} & \abinit, RLCCSDT & & & 42.87 \\ % verified
Hol10 \cite{Holmgren2010} & direct $\arb$ and $\ak$ meas. & & 47.24(44) & 43.06(36) \\
Hol10 \cite{Holmgren2010,Ekstrom1995} & ratio calibrated with $\alpha_{\text{Na}}$ & & 47.24(21) & 43.06(21) \\
Por10 \cite{Porsev2010,Rafac1998,Derevianko2001} & \abinit $D_{1/2}^2$ + $\rcs$ meas. + $\alpha_r$ & 59.42(3) & & \\
Saf11 \cite{Safronova2011} & \abinit, RCCSD & & 47.72(59) & \\ % gussed
Nan12 \cite{Nandy2012} & \abinit, RCCSDT & & & 43.05(15) \\ % guessed
Bor13 \cite{Borschevsky2013} & \abinit, RCCSDT & 59.13 & & \\ % guessed
Saf13 \cite{Safronova2013} & \abinit, RLCCSDT & & & 43.03(9) \\ % guessed
Sah13 \cite{Sahoo2013} & \abinit, RCCSDT & & & 42.94(9) \\ % guessed
Jai13 \cite{Jiang2013} & semi-empirical & & & 42.98 \\
Yon14 \cite{Y.-B.2014} & semi-empirical & 58.72 & 47.07 & 42.94 \\
Pat15 \cite{Patterson2015,Rafac1998,Derevianko2001}  & $\tau_{3/2}$ meas. + $\rcs$ meas. + $\alpha_r$ & 59.21(13) & & \\
This work & direct meas. & $\polCsSysOnly$ & $\polRbSysOnly$ & $\polKSysOnly$ \\
\hline\hline
\end{tabular}
\end{center}
\end{table*}
\endgroup

%CICP: configuration interaction with semi-empirical core potential
%RCCSDT: coupled cluster method with single and double excitations of Dirac-Fock wavefunctions with partial triple contributions
%RLCCSD: linearized coupled cluster method with single and double excitations of Dirac-Fock wavefunctions
%RCCSDT: linearized coupled cluster method with single and double excitations of Dirac-Fock wavefunctions with partial triple contributions

%CCSD(T): coupled cluster with single, double, and perturbative triple excitations
%SD: single double, in which single and double excitations of DF (Dirac-Fock) wavefunctions are iterated to all orders

We can also report a polarizability using $\rcs$ \cite{Rafac1998} in conjunction with Porsev \etal's 2010 calculation of $D_{1/2,\mathrm{Cs}}^2 = 20.334$ (in atomic units) \cite{Porsev2010}.
We can write $\alpha_i$ in terms of the electric dipole transition matrix elements as
\begin{align}
	\alpha_i = \frac{e^2}{12 \pi \epsilon_0 a_0^4} \sum_{j\neq i}	
	\frac{\left|\brakett{i}{\hat{D}}{j}\right|^2}{\hbar\omega_{ij}}	
	+ \alpha_r
	\label{polFromMatrixElements}
\end{align}
where $a_0$ is the Bohr radius. 
As before, we only explicitly consider the $ns_{1/2}-np_{1/2}$ and $ns_{1/2}-np_{3/2}$ matrix elements, 
where $ns_{1/2}$ is the ground state.
We abbreviate the matrix elements associated with the principal transitions to $D_{1/2}$ and $D_{3/2}$.

In 2002, Derevianko and Porsev demonstrated a method for obtaining values of $D_{1/2,\mathrm{Cs}}^2$ and $D_{3/2,\mathrm{Cs}}^2$ from Cs van der Waals $C_6$ coefficients \cite{Derevianko2001} and $\rcs$ \cite{Rafac1998}. \figref{comparisonsMisc} includes $\alpha$ values derived using experimental Cs $C_6$ measurements in conjunction with that method \cite{Leo2000,Chin2004}.

\subsection{Other atomic properties derived from our polarizability measurements} \label{sectionDerivedProperties}

Finally, we use our polarizability measurements to report matrix elements, lifetimes, oscillator strengths, line strengths, and van der Waals $C_6$ coefficients.
In these calculations, we use residual polarizabilities $\alpha_r$ and matrix element ratios $R$ from Table \ref{tableOmegaRes}.
To report matrix elements and lifetimes, we use \eqnref{polFromMatrixElements} and \eqnref{polFromLifetimes}. $\alpha_i$ is given in terms of oscillator strengths $f_{ij}$ as 
\begin{align}
	\alpha_i = \frac{e^2}{4 \pi \epsilon_0 m}
	\sum_{j \neq i}
	\frac{f_{ij}}{w_{ij}^2}
	+ \alpha_r
	\label{polFromOscStr}
\end{align}
where $m$ is the electron mass. 
$\alpha_i$ is also given in terms of line strengths $S_{ji}$ as
\begin{align}
	\alpha_i = \frac{1}{6\pi\epsilon_0\hbar} 
	\sum_{j \neq i} 
	\frac{S_{ji}}{g_i\omega_{ij}}
	+ \alpha_r
	\label{polFromLineStr}
\end{align}

$C_6$ can be expressed in terms of dynamic polarizability as
\begin{align}
	C_6 = \frac{3}{\pi} \int_0^{\infty} \alpha(i\omega)^2 d\omega
	\label{C6fromPol}
\end{align}
Derevianko \etal's 2010 work tabulates values of $\alpha(i\omega)$ for Cs, Rb, and K atoms among others \cite{Derevianko2010}. To derive $C_6$ values from our $\alpha(0)$ measurements, we modify Derevianko \etal's values of $\alpha(i\omega)$ to get
\begin{align}
	\alpha_{\mathrm{new}}(i\omega) =
	\tilde{\alpha}(i\omega) -
	\left[\alpha_{\textrm{this work}}(0) - \tilde{\alpha}(0)\right]
	\frac{\alpha_p(i\omega)}{\alpha_p(0)}
	\label{alphaNew}
\end{align}
where $\tilde{\alpha}(i\omega)$ and $\tilde{\alpha}(0)$ refer to values tabulated by Derevianko \etalspace
In the above equation (\ref{alphaNew}), $\alpha_p(i\omega)$ is the contribution to $\alpha(i\omega)$ by the principle transitions. The ratio $\alpha_p(i\omega)/\alpha_p(0)$ is given by
\begin{align}
	\frac{\alpha_p(i\omega)}{\alpha_p(0)} = \frac
	{
		\frac{1}{\omega_{1/2}^2+\omega^2} + \frac{R\frac{\omega_{1/2}}{\omega_{3/2}}}{\omega_{3/2}^2+\omega^2}
	}
	{
		\frac{1}{\omega_{1/2}^2} + R\frac{\omega_{1/2}}{\omega^3_{3/2}}
	}
	\label{polPRatio}
\end{align}

Predictions of the parity-non-conserving amplitude, $E_{\mathrm{PNC}}$, in Cs depends heavily on $D_{1/2}$.
We note that our Cs $D_{1/2}$ value is consistent with the theoretical Cs $D_{1/2}$ calculated by Porsev \etalspace in 2010 for the purpose of interpreting PNC data as a test of the standard model \cite{Porsev2010}.

Finally, we use our measurements together with recent measurements of Cs, Rb, and K $\alpha_{np_{1/2}} - \alpha_{ns_{1/2}}$ \cite{Hunter1991,Miller1994} to report excited state polarizabilities $\alpha_{np_{1/2}}$ with better than 0.08\% uncertainty. 
These results are shown in Table \ref{tablePolExcited} and serve as benchmark tests for calculations of dipole transition matrix elements for $p-d$ transitions.

\section{Outlook} \label{sectionOutlook}

We are currently exploring ways to measure the polarizability of Li and metastable He, the polarizabilities of which can be accurately calculated. By measuring $\acs/\alpha_{\mathrm{He*}}$ or $\acs/\alpha_{\mathrm{Li}}$, we could report $\acs$ with precision comparable to that of the ratios reported here for the benefit of PNC research. Such a measurement would also act as a calibration of the measurements presented in this work, because it would be independent of systematic errors that may affect our direct measurements.

We are also exploring electron-impact ionization schemes for atom detection, which would allow us to detect most atoms and molecules. Our Langmuir-Taylor detector only allows us to detect alkali metals and some alkaline-Earth metals \cite{Delhuille2002}. Installing an electron-impact ionization detector would allow us to broaden the scope of atom interferometry as a precision measurement tool. 

This work is supported by NSF Grant No. 1306308 and a NIST PMG. M.D.G. and R.T. are grateful for NSF GRFP Grant No. DGE-1143953 for support. 

%\bibliography{references}
%\bibliography{libraryManual}
\bibliography{library}

\end{document}